\documentclass[12pt,preprint]{elsarticle}
\usepackage[utf8]{inputenc}
\usepackage[T1]{fontenc}
\usepackage{hyperref,lineno,natbib,amsmath,amssymb,graphicx,multirow,algorithmic,color,fullpage}
\usepackage[expansion=false]{microtype}
\usepackage[skip=4pt]{caption}
\usepackage[title]{appendix}
\usepackage{longtable}
\usepackage{graphicx}
\usepackage{array}
\usepackage{siunitx}
\usepackage{subfigure}
\usepackage{bm}
\usepackage{placeins}
\newcommand{\vect}[1]{\mathbf{#1}}
\usepackage{tikz}

\usepackage[margin=1in]{geometry}
\usepackage{booktabs}
\usepackage{amsmath}
\usepackage{graphicx}

\usepackage{amssymb, amsmath, amsfonts}
\usepackage{outlines}

\usepackage{enumitem}
\setlist[enumerate,1]{label=\textit{\Roman*}}
\setlist[enumerate,2]{label=\textit{\roman*}}
\setlist[enumerate,3]{label=\textit{\alph*}}

\journal{Computer Physics Communications}
\usepackage{xcolor}
\usepackage{listings}

\usepackage{float}

\colorlet{mygray}{black!30}
\colorlet{mygreen}{green!60!blue}
\colorlet{mymauve}{red!60!blue}

\definecolor{webgreen}{rgb}{0,.35,0}
\definecolor{webbrown}{rgb}{.6,0,0}
\definecolor{RoyalBlue}{rgb}{0,0,0.9}
\definecolor{mywhite}{rgb}{1.0,1.0,1.0}

\definecolor{purp}{rgb}{0.4,0.2,0.8}

\hypersetup{
   colorlinks=true, linktocpage=true, pdfstartpage=3, pdfstartview=FitV,
   breaklinks=true, pdfpagemode=UseNone, pageanchor=true, pdfpagemode=UseOutlines,
   plainpages=false, bookmarksnumbered, bookmarksopen=true, bookmarksopenlevel=1,
   hypertexnames=true, pdfhighlight=/O,
   urlcolor=webbrown, linkcolor=RoyalBlue, citecolor=webgreen,
   pdfcreator={pdfLaTeX},
   pdfproducer={LaTeX with hyperref}
}

\newcommand{\R}{\mathbb{R}}

\newcommand{\co}[1]{\texttt{#1}}
\newcommand{\vpp}{\textsc{Voro++}}
\newcommand{\vplex}{\textsc{VoroPlex}}
\renewcommand{\vec}[1]{\mathbf{#1}}
\newcommand{\vp}{\vec{p}}
\newcommand{\vx}{\vec{x}}

\newcommand{\vc}{\vec{c}}
\newcommand{\trans}{\mathsf{T}}

\usepackage{xcolor}

\definecolor{colrev}{rgb}{0.9,0,0}

\usepackage{algorithm,algorithmic}

\newlength\myindent

\newlist{steps}{enumerate}{1}
\setlist[steps, 1]{label = \textit{Step \arabic*}}

\usepackage{xcolor}

\begin{document}
\begin{frontmatter}

  \title{\vplex: A Python Framework for Multi-Threaded and Differentiable
    Voronoi Tessellation}

\author[UCLA,Yale]{Peihang Lin\fnref{equal}}
\author[UCLA]{Shanmei Wanyan\fnref{equal}}
\author[UCLA]{Mucheng Zhu\fnref{equal}}
\author[UCLA]{Ying Jiang}
\author[UW,LBL]{Chris H. Rycroft}
\author[UCLA]{Chenfanfu Jiang}
\author[UCLA,UW]{Jiayin Lu\corref{cor}}
\ead{jiayin.kay.lu@gmail.com}
\address[UCLA]{Department of Mathematics, University of California, Los Angeles,
Los Angeles, CA 90095, United States}
\address[Yale]{Department of Statistics and Data Science,
Yale University, New Haven, CT 06511, United States}
\address[UW]{Department of Mathematics, University of Wisconsin--Madison,
Madison, WI 53706, United States}
\address[LBL]{Mathematics Group, Lawrence Berkeley National Laboratory,
Berkeley, CA 94720, United States}

\fntext[equal]{These authors contributed equally to this work and are listed alphabetically by surname.}
\cortext[cor]{Corresponding author.}

\begin{abstract}
Voronoi tessellations are widely used in computational physics, for example to
measure local structure in particle systems and to generate meshes. We present
\vplex, an open-source Python package for computing Voronoi tessellations in two
and three dimensions. \vplex{} is built on the multi-threaded version of the C++
library \vpp, so that large tessellations can be computed in parallel. The
package also makes the tessellation differentiable. Each Voronoi vertex is
labeled by the generators that define it, and \vplex{} uses these labels to
compute the derivative of the vertex with respect to the positions of these
generators. PyTorch can therefore compute gradients of quantities such as cell
areas, volumes, and interface areas with respect to the generator positions, so
that these quantities can be optimized using standard gradient-based methods.
The package also includes tools for extracting interfaces between groups of
cells and for exporting meshes. For one million points using 28 threads,
\vplex{} is 46 times faster than SciPy in 2D and 39 times faster in 3D. We
demonstrate the package with three applications: centroidal Voronoi
tessellations with a variable density, meshing of 2D shapes defined by signed
distance functions, and the design of foams with a preferred cell orientation.
\end{abstract}

\begin{keyword}
Voronoi tessellation \sep differentiable Voronoi tessellation \sep computational
geometry \sep parallel computing \sep OpenMP \sep automatic differentiation \sep
PyTorch \sep centroidal Voronoi tessellation
\end{keyword}
\end{frontmatter}

\section*{Program summary}

\noindent \textit{Program title:} \vplex{} \\ \vspace{-0.3em}

\noindent \textit{Developer's repository link:}
\url{https://github.com/rycroft-group/voroplex} \\  \vspace{-0.3em}

\noindent \textit{Licensing provisions:} BSD 3-clause \\  \vspace{-0.3em}

\noindent \textit{Programming language:} Python, C++, Cython  \\ \vspace{-0.3em}

\noindent \textit{External routines/libraries:}
OpenMP, multi-threaded \vpp, NumPy, PyTorch, VTK (optional) \\ \vspace{-0.3em}

\noindent \textit{Nature of problem:} Voronoi tessellations are used in many
areas of computational physics, but the fastest libraries for computing them are
written in C++. Existing Python packages for Voronoi tessellations run on a
single thread and can be slow for large systems. In addition, they do not
provide the tessellation in a differentiable form. This makes it difficult to
use gradient-based optimization to adjust the generator positions based on
quantities such as cell volumes or interface areas.\\ \vspace{-0.3em}

\noindent \textit{Solution method:} \vplex{} provides a Python interface,
written in Cython, to the multi-threaded version of \vpp, and computes Voronoi
tessellations in parallel in two and three dimensions. The result is returned in
NumPy arrays, in one of three representations: the vertices and faces of each
cell, as computed by \vpp; the same data, with each vertex also labeled by the
IDs of the generators that define it; or only these labels, which are used to
build differentiable meshes in which each vertex is shared between cells. To
make the tessellation differentiable, \vplex{} returns the vertices of each cell
as a PyTorch tensor. When PyTorch computes the gradient of a quantity with
respect to these vertices, \vplex{} converts it into a gradient with respect to
the generator positions. This uses the fact that each vertex is at the same
distance from all of the generators that define it, or lies on the boundary of
the domain. \vplex{} also provides tools for extracting interfaces, analyzing
cells, and exporting results for visualization.\\ \vspace{-0.3em}

\section{Introduction}
\label{sec:intro}

\subsection{Voronoi tessellations and their uses}

Many problems in computational physics require a way to divide space into
regions around a set of points. The Voronoi tessellation provides a natural way
to do this, and it is widely used to study the local structure of particle
systems, to generate meshes, and to optimize the placement of points.

\begin{figure}[!t]
  \centering
  \includegraphics[width=1\textwidth]{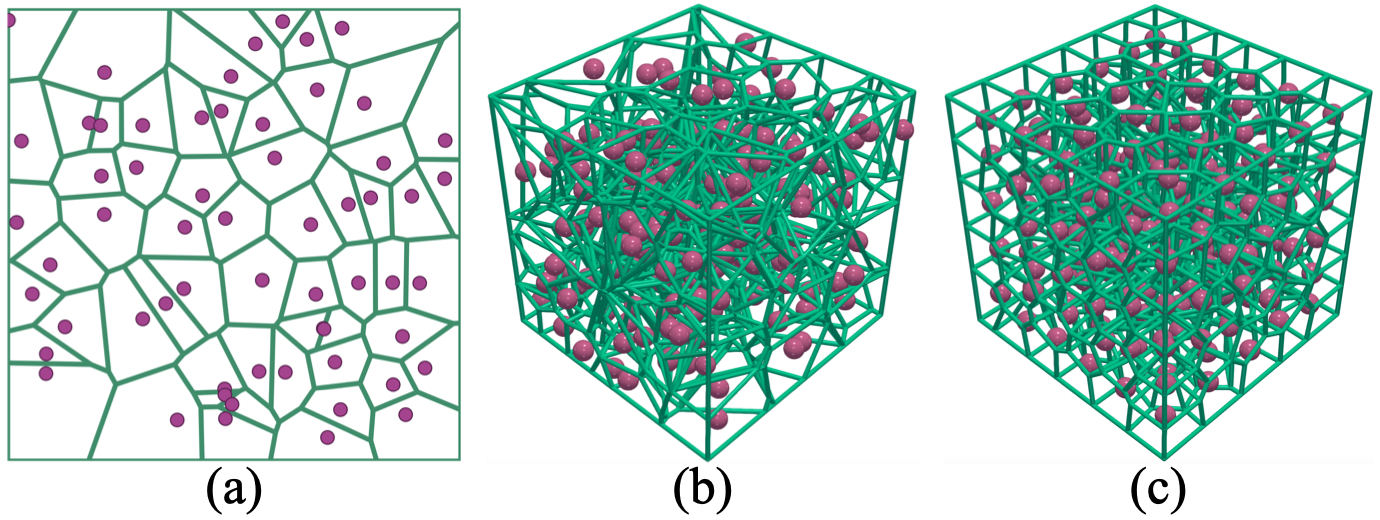}
  \captionsetup{skip=0.8em}
  \caption{Voronoi tessellations with random generators in (a) 2D, consisting of
  polygonal cells, and (b) 3D, consisting of polyhedral cells. (c) Centroidal
  Voronoi tessellation in 3D with a uniform density field.}
	\label{fig:voronoi-examples}
\end{figure}

Let $\Omega \subset \R^d$ be a computational domain, with $d=2$ or $3$, and let
$\{\vp_i\}_{i=1}^N \subset \Omega$ be a set of generator points. The Voronoi
cell associated with generator $\vp_i$ is defined as
\begin{equation}
	V_i =
	\left\{
	\vx \in \Omega :
	\|\vx-\vp_i\| \leq \|\vx-\vp_j\|,
	\quad \forall j \neq i
	\right\}.
	\label{eq:voronoi-cell}
\end{equation}
In other words, $V_i$ is the set of points that are at least as close to $\vp_i$
as to any other generator. The collection $\{V_i\}_{i=1}^N$ forms a Voronoi
tessellation of the domain~\cite{Aurenhammer91,okabe09}. As shown in
Fig.~\ref{fig:voronoi-examples}(a,b), the Voronoi cells are polygons in two
dimensions and polyhedra in three dimensions. Neighboring cells meet along edges
in two dimensions and along polygonal faces in three dimensions.

Voronoi tessellations are used to study systems across a wide range of length
scales~\cite{lazar2022}. At the atomic scale, the Voronoi cell of an atom
gives a natural measure of its volume and local environment, which has been
used to study the structure of interfaces between phases~\cite{yang15,sun15}.
Polycrystalline materials have long been modeled by representing each grain as
a Voronoi cell~\cite{kumar94,fan04,benedetti13}, and these models have been
used to simulate cracking~\cite{gulizzi18,benedetti18}. In granular materials
and other packings, Voronoi cells are used to measure the local packing density
and structure, in both experiments~\cite{panaitescu10} and
simulations~\cite{yang2002,rycroft10,PHUA2023,rieser16}. They have also been
used to relate the local structure of disordered networks to bulk properties
such as electrical resistance~\cite{wang2026resistance}. At larger scales,
Voronoi tessellations have been used to measure the local concentration of
phytoplankton in turbulent flow~\cite{durham13}, to estimate the density of
snow aggregates~\cite{honeyager16}, to allocate resources between geographical
regions~\cite{feng2018heterogeneous,Li2025}, and to identify voids in the
large-scale structure of the universe~\cite{ruiz15}.

The Voronoi tessellation is the dual of the Delaunay triangulation, which
connects each pair of generators whose cells share a boundary. Delaunay
triangulations are the basis of many methods for generating finite element
meshes~\cite{shewchuk96,Shewchuk98TetRefine,cheng2012,claes_fem,Leyssens2024}.
The Voronoi cells themselves can also be used as the control volumes in finite
volume simulations~\cite{moukalled2016}. Since the face between two cells is
perpendicular to the line joining their generators, the flux between the cells
is simple to compute. Voronoi meshes have been used in this way to model flow
and transport in the subsurface, including in nuclear waste
repositories~\cite{freeman14,blanco-martin16}, and to simulate radiative
transfer in astrophysics~\cite{camps13,camps15}.

A centroidal Voronoi tessellation (CVT) is a Voronoi tessellation in which
each generator lies at the centroid of its own cell, as shown in
Fig.~\ref{fig:voronoi-examples}(c). The cells of a CVT are compact and roughly
equal in size, and although an individual cell may have preferred directions,
there is no preferred orientation on average. A density function $\rho(\vx)$
can be used to weight the calculation of the centroids, so that the generators
cluster where $\rho$ is large. This allows the size of the cells to be
controlled in different parts of the domain. CVTs have many
applications~\cite{du1999centroidal}, including mesh generation~\cite{du02},
the construction of grids on the sphere for climate
modeling~\cite{jacobsen13}, and image compression~\cite{du2006cvt}. In
addition, generalizations of CVTs have been used to solve optimal location
problems~\cite{bourne15}.

\subsection{Existing software for Voronoi tessellations}

Many algorithms for computing Voronoi tessellations have been
developed~\cite{Aurenhammer91,okabe09}. A widely used approach is to compute the
Delaunay triangulation and then construct the Voronoi tessellation from it. This
approach is used by the Qhull library~\cite{qhull_lib}, which is available in
Python through \co{scipy.spatial.Voronoi}~\cite{virtanen2020scipy}, and by the
C++ libraries CGAL~\cite{cgal:eb-23b} and Geogram~\cite{geogram}. Many of these
algorithms build the triangulation incrementally, by adding the generators one
at a time~\cite{green78,lee80}. With these approaches, the whole triangulation
must be stored, which can require a large amount of memory for large
three-dimensional systems, and quantities for individual cells, such as their
volumes, must be assembled in a separate step. Fortune's sweep-line
algorithm~\cite{fortune86} is very efficient in two dimensions, but it is
inherently sequential, and it has no practical extension to three dimensions. It
is used by the C++ library
Boost.Polygon~\cite{simonson2009geometry,sydorchuk_boost_voronoi}, which is
available in Python through pyvoronoi~\cite{ancelin2025pyvoronoi}. Grid-based
methods use the fast marching method~\cite{sethian99,kimmel98} to find the
nearest generator to each grid point. These methods are more expensive and only
approximate the cells, but they can use other ways of measuring distance, such
as the distance along a curved surface~\cite{kimmel01a}.

\begin{figure}[!t]
  \begin{center}
    \includegraphics[width=1\textwidth]{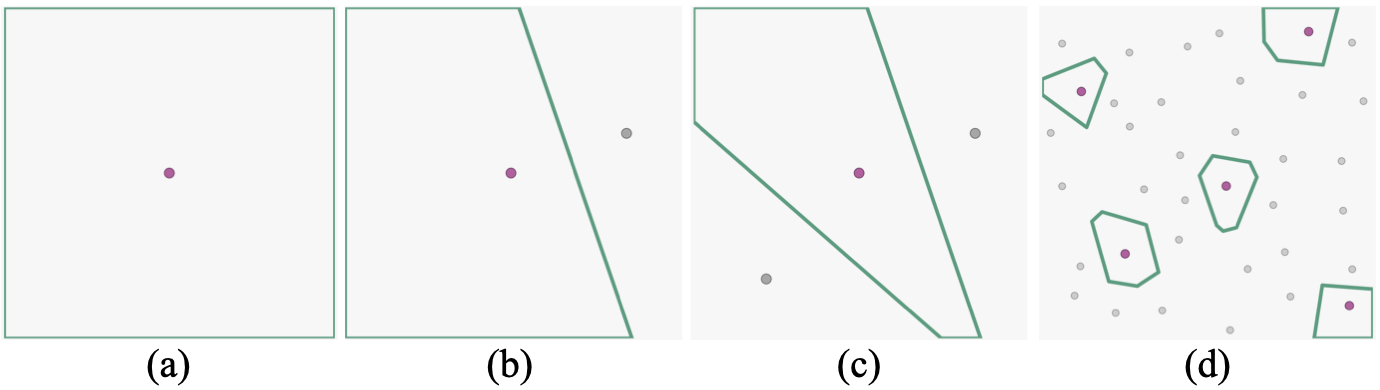}
  \end{center}
  \caption{Computing a Voronoi cell separately. (a) The cell of a generator
  (purple) starts as the whole domain. (b) The cell is cut by the perpendicular
  bisector between the generator and a neighboring generator (gray). (c)
  Further cuts by other neighbors produce the final cell. (d) Since each cell
  is computed separately, many cells can be computed in parallel.}
  \label{fig:cell_based_approach}
\end{figure}

The approach used in this work is to compute each Voronoi cell separately. The
cell of a generator starts as a large shape, such as the domain, and it is then
cut by the perpendicular bisector planes between the generator and each of its
nearby neighbors, as shown in Fig.~\ref{fig:cell_based_approach}. Early
algorithms constructed the cell of each point in the plane in this
way~\cite{rhynsburger73}, and used a grid to find the nearby points quickly, so
that the whole tessellation could be computed in linear expected
time~\cite{bentley80}. The approach was also used to simulate random Voronoi
cells in two and three dimensions, by generating points in order of increasing
distance until no further point could affect the cell~\cite{boots83,quine84}. It
is used by the \vpp{} library~\cite{rycroft09c}, and it is well suited to
particle systems, where the quantities of interest are usually properties of
individual cells.

Since each cell is computed separately using floating-point arithmetic,
neighboring cells do not always agree exactly about their shared topology. For
example, the cells of four generators at the corners of a square should all
meet at a single vertex. However, rounding error may give one cell a very short
extra edge shared with the diagonally opposite generator, while the cell of
the opposite generator does not have this edge. In three dimensions, the same
effect creates very small extra faces. This is common for crystalline
arrangements of particles, where many generators lie on common circles or
spheres, and it can also occur by chance for random arrangements. Quantities
that depend continuously on the vertex positions, such as the volumes, surface
areas, and centroids of the cells, are not affected. However, quantities that
describe the topology of a cell, such as its number of faces and its list of
neighbors, can be affected.

The \vpp{} library computes each cell to within a small tolerance, and it
deliberately does not attempt to make the topology exactly consistent between
cells. Exact consistency can be achieved using adaptive-precision
arithmetic~\cite{shewchuk97} or arbitrary-precision
libraries~\cite{gmp,fousse07}, but this adds considerable complexity. Exact
topology is important for some mathematical problems. However, for experimental
or simulation data, a measurement that depends on topological differences at the
level of rounding error, around $10^{-16}$ in double precision, is difficult to
interpret, since the errors in the data are usually far larger. For example, Sun
\textit{et al.}~\cite{sun15} removed the tiny Voronoi faces created by thermal
vibrations in molecular dynamics simulations before identifying the neighbors of
each atom. Lazar \textit{et al.}\ have developed a more complete framework, in
which the space of possible Voronoi cells is divided into regions with different
topologies~\cite{lazar12,lazar13,lazar15}. Some crystal structures, such as
face-centered cubic, lie on the boundaries between regions, so small
perturbations produce a family of different topologies. By analyzing the
statistics of the topologies over many cells, rather than the topology of any
single cell, this approach is insensitive to rounding error. It is implemented
in the VoroTop software~\cite{lazar17}, which has recently been extended to two
dimensions and to use the multi-threaded version of \vpp{} discussed
below~\cite{lazar24}.

Computing the cells separately also makes the approach naturally parallel.
The multi-threaded version of \vpp{}~\cite{lu2022} uses OpenMP~\cite{dagum98}
to compute the cells on several threads at once. Any of the standard OpenMP
scheduling options can be used. For example, with dynamic scheduling, each
thread starts a new cell as soon as it finishes its previous one, which
balances the work well when the cells take different amounts of time to
compute. A similar approach has been used to compute Voronoi cells on a
GPU~\cite{ray2018meshless}, and parallel algorithms have also been developed
for the Delaunay-based approach~\cite{jacobsen13}.

Several Python packages provide access to \vpp, including pyvoro~\cite{pyvoro},
tess~\cite{tess}, and freud~\cite{ramasubramani2020freud}, a toolkit for
analyzing particle simulations in periodic boxes. However, all of these packages
use the original, serial version of \vpp, and there is currently no Python
package for computing Voronoi cells in parallel with the multi-threaded version.

\subsection{Differentiable Voronoi tessellations}

Many applications require finding generator positions that minimize an
objective function that depends on the geometry of the Voronoi cells. The
classic example is the CVT, which minimizes the energy
\begin{equation}
  E = \sum_{i=1}^N \int_{V_i} \|\vx-\vp_i\|^2 \, d\vx .
  \label{eq:cvt-uniform}
\end{equation}
The gradient of this energy has a simple form: the derivative with respect to
$\vp_i$ is $2A_i(\vp_i-\vc_i)$, where $A_i$ and $\vc_i$ are the area (or
volume) and the centroid of cell $V_i$~\cite{du1999centroidal}. This is the
basis of Lloyd's algorithm~\cite{lloyd82}, which repeatedly moves each
generator to the centroid of its cell, and of faster quasi-Newton
methods~\cite{Liu2009CVTsmooth}.

For most other objective functions, however, the gradient does not have a simple
form. Moving one generator changes the shape of all the neighboring cells, and
working out the resulting gradient by hand is tedious and prone to errors.
Automatic differentiation, available in libraries such as
PyTorch~\cite{paszke2017automatic}, avoids this. The user writes code to
evaluate the objective function, and the library computes its gradient
automatically. To use automatic differentiation, the Voronoi geometry must be
written as a differentiable function of the generator positions. This is
possible because each Voronoi vertex is the solution of a small linear system
involving the generators that define it, and its position therefore changes
smoothly as the generators move. However, this only holds while the topology of
the tessellation stays the same. As the generators move, cells gain or lose
neighbors, and the set of generators that defines each vertex changes. The
solution of the original linear system is then no longer a vertex of the
tessellation, because some other generator is now closer to it. The tessellation
must therefore be recomputed as the generators move, so that the derivative of
each vertex is always computed from the correct set of generators.

Several recent studies have used differentiable Voronoi tessellations. Numerow
\textit{et al.}~\cite{Numerow2024DiffVoronoi} used them to simulate
cell-based mechanical systems, and Maruani \textit{et
al.}~\cite{maruani2023voromesh} used them to learn watertight surface
meshes. These methods were developed for specific
applications, and they do not include fast, parallel computation of
the tessellation.

\subsection{This work}

In this paper we present \vplex, a Python package that combines
parallel computation of Voronoi tessellations with differentiable geometry.
Its main features are:

\begin{enumerate}
	\item A Python interface to the multi-threaded \vpp{} library, for
	computing Voronoi tessellations in two and three dimensions, and radical
	tessellations of particles with different sizes. The results are returned
	as NumPy arrays.

	\item Three representations of the result: the vertices and faces of each cell,
	as computed by \vpp; the same data, with each vertex also labeled by the
	integer IDs of the generators that define it; and these labels alone, which are
	used to build differentiable meshes in which each vertex is shared between
	cells.

	\item Differentiable vertices. \vplex{} returns the vertices of each cell as a
	PyTorch tensor, and computes the derivative of each vertex with respect to the
	positions of the generators that define it. The gradient of any quantity
	computed from the vertices can therefore be found with respect to the generator
	positions, on either the CPU or the GPU.

	\item Tools for extracting interfaces between groups of cells, analyzing
	cell geometry, and exporting results for visualization.
\end{enumerate}

The rest of this paper is organized as follows. Section~\ref{sec:design}
describes the design of the software, and Section~\ref{sec:examples} gives
examples of how to use it. Section~\ref{sec:performance} compares its
performance with other Python packages and examines its parallel scaling.
Section~\ref{sec:applications} presents three applications of differentiable
optimization: CVTs with a variable density, meshing of two-dimensional shapes
defined by signed distance functions, and the design of foams with a
preferred cell orientation. Section~\ref{sec:conclusion} gives our
conclusions.

\section{Software design}
\label{sec:design}

\subsection{Overview}

\vplex{} has three layers. The multi-threaded \vpp{} library, written in C++,
computes the tessellation. A layer written in Cython passes the data between
C++ and Python. The Python layer provides the user interface, and it connects
the Voronoi vertices to PyTorch's automatic differentiation. A typical
calculation has the following steps:
\begin{enumerate}[label=(\arabic*)]
\item The user provides the generator positions as a NumPy
  array~\cite{harris2020array}.
\item \vpp{} computes the Voronoi cells in parallel on the CPU. For each vertex
  of each cell, \vplex{} records its coordinates and, if needed, its signature,
  which lists the generators that define it.
\item The vertices of every cell are returned as a PyTorch tensor.
\item The user evaluates an objective function from the vertices using
  PyTorch.
\item To compute the gradient of the objective function with respect to the
  generator positions, PyTorch differentiates the objective function, and
  \vplex{} supplies the derivative of each vertex with respect to its
  generators, which it computes analytically.
\end{enumerate}
Steps (1) and (2) are needed for any use of the package, while steps (3)--(5)
are only needed when the tessellation must be differentiated. The main
functions are listed in Table~\ref{tab:functions}. Most functions have separate
versions for two and three dimensions. The data for all of the cells are stored
together in a small number of flat arrays, rather than in a separate Python
object for each cell, and all calculations on these arrays are vectorized, so
that there are no loops over cells or vertices in Python. The benefits of this
design are discussed in Section~\ref{sec:performance}.

\begin{table}[!t]
\centering
\begin{tabular}{p{0.44\textwidth}p{0.5\textwidth}}
\toprule
Function & Purpose \\
\midrule
\co{compute\_voronoi\_2d}, \co{compute\_voronoi\_3d} &
  Compute the tessellation, either standard or radical, and return the
  result as a \co{VoronoiResult} object. \\
\co{compute\_properties} &
  Compute further properties of the cells for an existing result. \\
\co{cell\_vertices} &
  Return the vertices of every cell as a differentiable PyTorch tensor. \\
\co{cell\_fan\_simplices} &
  Split every cell into triangles (2D) or tetrahedra (3D), each made from the
  cell's generator and one edge or part of a face. \\
\midrule
\co{compute\_unique\_vertices} &
  Compute the positions of the unique vertices, each shared between cells. \\
\co{cell\_polytopes} &
  List the vertices of each cell in order (2D), or its faces (3D), using the
  indices of the unique vertices. \\
\co{unique\_facets} &
  List each edge (2D) or face (3D) of the tessellation once. \\
\co{surface\_mesh} &
  Extract the mesh of edges (2D) or faces (3D) on the interface between inner
  and outer cells, and on the boundary of the domain. \\
\midrule
\co{analysis.compute\_polygon\_geometry},
  \co{analysis.compute\_surface\_geometry} &
  Compute the areas, perimeters, normals, and neighbors of cells (2D) or faces
  (3D), in a differentiable form. \\
\co{export\_vtu}, \co{export\_npz}, \co{load\_npz} &
  Export a tessellation for visualization in ParaView, and save or load a
  result. \\
\bottomrule
\end{tabular}
\caption{The main functions in \vplex. The functions after the first two work
in both 2D and 3D, and find the dimension from the result. Those in the middle
group build meshes in which each vertex is shared between the cells that meet
there. \co{compute\_unique\_vertices} requires the topology representation;
the others also work with the default geometry representation, except in
periodic domains.}
\label{tab:functions}
\end{table}

\subsection{Computing the tessellation}
\label{sec:computing}

The functions \co{compute\_voronoi\_2d} and \co{compute\_voronoi\_3d} take an
$N\times d$ array of generator positions, where $d$ is the number of
dimensions. The domain is a rectangular box. Its limits can be given by the
user, and otherwise the box extends 5\% beyond the generators in each
direction. Each direction can either be periodic, or bounded at both ends. The
cells are computed by \vpp{} using the cell-based method described
in Section~\ref{sec:intro} and shown in Fig.~\ref{fig:cell_based_approach}.
\vplex{} divides the cells between threads using OpenMP's dynamic scheduling,
since the time needed to compute a cell varies, for example between cells in
dense and sparse regions~\cite{lu2022}. The number of threads can be set by the
user, or chosen automatically.

Two optional arguments allow the generators to be divided into groups. First,
all $N$ generators are used when computing the cells, but cells are only
computed for the first $n_\text{compute}$ generators. The remaining generators
affect the shapes of the computed cells, but they do not have cells of their
own. This is useful, for example, when extra generators are placed outside a
region only to define the shapes of the cells inside it. Second, the generators
can be divided into inner and outer sets, where the first $n_\text{inner}$
generators are inner. \vplex{} then records which inner cells touch an outer
cell, and which touch the boundary of the domain. The interface between the
inner and outer cells can also be extracted as a mesh, as described in
Section~\ref{sec:interfaces}. In both cases, the generators must be ordered so
that those in the first group come first.

The generators can also be given radii $r_i$, with the optional \co{radii}
argument. \vplex{} then computes the radical Voronoi tessellation, using the
corresponding \vpp{} classes. In this tessellation, the cell of generator $i$
is the set of points $\vx$ for which
\begin{equation}
  \|\vx-\vp_i\|^2-r_i^2 < \|\vx-\vp_j\|^2-r_j^2
  \qquad \text{for all } j\ne i.
  \label{eq:radical}
\end{equation}
The same tessellation is known by several names. The quantity
$\|\vx-\vp_i\|^2-r_i^2$ is called the power of the point $\vx$ with respect to
the sphere of radius $r_i$ centered on $\vp_i$, and in computational geometry
the tessellation is usually called the power diagram~\cite{aurenhammer87}. For
a point outside the sphere, the square root of the power is the length of a
tangent line from the point to the sphere, which is the distance used in
Laguerre geometry, and so the tessellation is also called the Voronoi diagram
in the Laguerre geometry~\cite{imai85}, the Laguerre
tessellation~\cite{lautensack08}, or the Laguerre--Voronoi
diagram~\cite{pekmezi24}. In the physical sciences, it is
often called the radical tessellation~\cite{marck96,pinheiro13a}, since the
face between two cells lies on the radical plane of the two spheres (the
radical axis of two circles in 2D), where their powers are equal.

The radical tessellation is used for systems of particles with different
sizes, where the standard Voronoi tessellation can cut through the larger
particles. With the radical tessellation, the face between two particles that
do not overlap lies in the gap between them, and the face between two
overlapping particles passes through the circle (in 3D) or the pair of points
(in 2D) where their surfaces intersect~\cite{marck96}. Applications include
the void space in packings of spheres~\cite{marck96} and in porous
materials~\cite{pinheiro13a}, models of material structure based on
packings of spheres of different sizes~\cite{burtseva15}, models of
polycrystals~\cite{depriester19}, the assignment of electron density to atoms
in molecular simulations~\cite{brehm21}, and boundary conditions for
discrete element simulations~\cite{pekmezi24}.

The radical cells are still convex polygons or polyhedra, so \vpp{} computes
them with the same algorithm, and with equal radii they are the Voronoi
cells. There are two differences. A generator can lie outside its own cell,
and a cell can be empty, for example when a small particle lies inside larger
ones; an empty cell has zero area or volume, and no vertices. All three
representations can be used, and the derivatives of the vertices can also be
found with respect to the radii, as described in
Section~\ref{sec:differentiable}.

The result is returned as a \co{VoronoiResult} object. This contains the
requested properties of each cell, which \vpp{} computes during the parallel
loop over the cells. By default, this is the area (in 2D) or volume (in 3D) of
each cell. The \co{properties} argument can also request the perimeter or
surface area, the centroid, the numbers of faces, vertices, and edges, the total
length of the edges, the largest distance from the generator to a vertex, and
the second moment $\int_{V_i}\|\vx-\vp_i\|^2\,d\vx$ and the corresponding
tensor, which we added to \vpp. Further properties can be computed later with
\co{compute\_properties}, which computes the tessellation again from the same
generators, storing only the new properties. The result also contains lists of
the cells that touch the interface or the boundary of the domain, and arrays
that describe the faces and neighbors of each cell. These arrays are read-only,
so that they cannot be changed by mistake.

\subsection{Three output representations}

The \co{representation} argument selects one of three representations for
the result:
\begin{itemize}
\item The \emph{geometry} representation stores the vertex coordinates, faces,
  and neighbors of each cell, exactly as computed by \vpp. It is the fastest
  option, and it is suitable when only the tessellation itself is needed, for
  example when analyzing the local structure of a particle simulation.
\item The \emph{per-cell} representation stores the same information, and in
  addition, the signature of each vertex of each cell, which lists the
  generators that define it. This representation is used for differentiation,
  as described in Section~\ref{sec:differentiable}. Each cell keeps its own
  copy of each of its vertices, so a vertex that is shared by several cells
  appears once for each of them. The cells therefore never need to be matched
  with each other.
\item The \emph{topology} representation stores the signatures but not the
  vertex coordinates. The copies of a vertex in different cells are matched
  using their signatures, so that each vertex is stored only once, and its
  position is recomputed in PyTorch from the generators that define it. A mesh
  in which the cells share their vertices is therefore a differentiable function
  of the generator positions. This is useful, for example, for the interface
  between two groups of cells, which can be extracted as a mesh, as described
  in Section~\ref{sec:interfaces}.
\end{itemize}
In all three representations, the tessellation is computed by \vpp{} on the
CPU. The representations are compared in Section~\ref{sec:performance}.
Alternatively, with \co{representation=None}, no information about the cells
themselves is stored, only their requested properties. This is useful when
only those properties are needed, as in Lloyd's algorithm for computing a CVT,
which only needs the centroids. For $10^6$ generators in 3D, it reduces the
memory used by the result from about 1.2\,kB per cell to 24 bytes per cell for
the centroids, and the computation is about 25\% faster.

\subsection{Vertex signatures}

\begin{figure}[!t]
  \centering
  \includegraphics[width=0.75\textwidth]{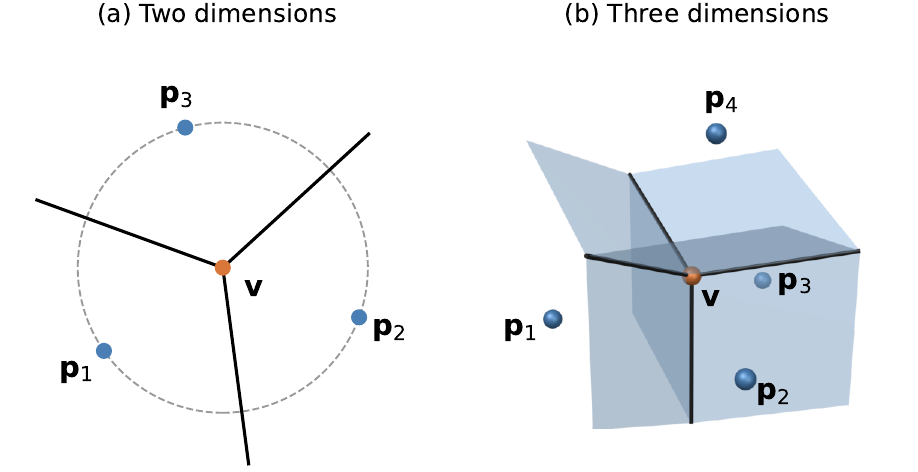}
  \captionsetup{skip=0.8em}
  \caption{A Voronoi vertex $\vect{v}$ (orange) is usually defined by (a) three
  generators in two dimensions, or (b) four generators in three dimensions. The
  generators are blue. The vertex is equidistant from its generators: in (a), it
  is the center of the circle (dashed) through them. In (a), three Voronoi edges
  meet at $\vect{v}$, each on the perpendicular bisector of two generators. In
  (b), four edges meet at $\vect{v}$, each equidistant from three generators,
  together with parts of the six faces (transparent) between pairs of
  generators. The IDs of the generators form the signature of the vertex.}
  \label{fig:signatures}
\end{figure}

In two dimensions, a vertex is usually defined by three generators, and it lies
at the point that is equidistant from all three. In three dimensions, a vertex
is usually defined by four generators. These cases are shown in
Fig.~\ref{fig:signatures}. As each cell is computed, \vplex{} records the
signature of each of its vertices. This is the sorted list of the IDs of the
cell's own generator and of the neighboring generators whose faces meet at the
vertex. Vertices on the boundary of the domain are defined partly by it, and so
each of the $2d$ faces of the box is given a negative ID, from $-1$ to $-2d$.
For example, a vertex where cell~7 meets cells~12 and~31 on the boundary
$x=x_\text{min}$ has the signature $[-1,7,12,31]$. If more than $d+1$ generators
are equidistant from a vertex, its signature has more than $d+1$ entries. The
signatures contain only the IDs of the generators, even in a periodic domain;
Section~\ref{sec:differentiable} describes how periodic images are handled.

The signatures are also used to build meshes in which each vertex is shared
between the cells that meet there. Each cell that shares a vertex produces the
same signature for it, so the unique vertices can be found by sorting the
signatures and removing duplicates, using exact integer comparisons rather than
comparing floating-point coordinates. For the geometry representation, which
does not store the signatures, they are first found from the neighbors of the
faces that meet at each vertex, and the position of each unique vertex is the
one computed by \vpp. For the topology representation, the position is computed
from the generators, using Eq.~\eqref{eq:bisector} below, so that it is
differentiable. Since the signatures do not record which periodic image of a
generator is meant, these meshes are not available in periodic domains. As
discussed in Section~\ref{sec:intro}, neighboring cells can occasionally
disagree about their shared topology when the generators are nearly degenerate.
The cells then produce different signatures for the same vertex, and the vertex
appears more than once in the list of unique vertices, so the mesh is not
exactly watertight at these points. This does not affect the differentiable
per-cell vertices, which never need to be matched between cells.

\subsection{Differentiable vertices}
\label{sec:differentiable}

The function \co{cell\_vertices} returns the vertices of every cell, cell by
cell, as a PyTorch tensor. Their coordinates are taken directly from \vpp, so
they do not need to be recomputed. The user then evaluates an objective
function $E$ from these vertices using PyTorch. For example, the function
\co{cell\_fan\_simplices} splits each cell into triangles in 2D, each made from
the cell's generator and one of its edges, or into tetrahedra in 3D, each made
from the generator and a triangle from one of its faces. Integrals over the
cells can then be computed as sums over these simplices, as for the CVT energy
in Section~\ref{sec:examples}. When the gradient of $E$ is needed, PyTorch
computes the derivative $\vect{g}=\partial E/\partial\vect{v}$ with respect to
each vertex $\vect{v}$, and \vplex{} converts these into derivatives with
respect to the generator positions, as follows.

Consider a vertex $\vect{v}$ in three dimensions with generators
$\vect{p}_1,\ldots,\vect{p}_4$, as in Fig.~\ref{fig:signatures}(b). Since the
vertex is equidistant from all four generators,
$\|\vect{v}-\vect{p}_j\|^2=\|\vect{v}-\vect{p}_1\|^2$ for $j=2,3,4$. The
quadratic terms $\|\vect{v}\|^2$ cancel on both sides, which leaves three
linear equations,
\begin{equation}
  2(\vect{p}_j - \vect{p}_1)\cdot\vect{v}
  = \|\vect{p}_j\|^2 - \|\vect{p}_1\|^2,
  \qquad j = 2,3,4.
  \label{eq:bisector}
\end{equation}
Each equation states that $\vect{v}$ lies on the perpendicular bisector plane
between $\vect{p}_1$ and $\vect{p}_j$. We write these equations as
$A\vect{v}=\vect{b}$, where the row of the $3\times3$ matrix $A$ for each $j$
is $2(\vect{p}_j-\vect{p}_1)^\trans$. In two dimensions, the same equations with
$j=2,3$ give a $2\times2$ system. For a vertex on the boundary of the domain,
one of these equations is replaced by the position of the boundary, such as
$v_x=x_\text{min}$.
If the generators move by small amounts $d\vect{p}_j$, then differentiating
Eq.~\eqref{eq:bisector} shows that the vertex moves by $d\vect{v}$, where
\begin{equation}
  A\,d\vect{v} = \vect{r}, \qquad
  r_j = 2(\vect{p}_j-\vect{v})\cdot d\vect{p}_j
      - 2(\vect{p}_1-\vect{v})\cdot d\vect{p}_1 .
  \label{eq:vertex-derivative}
\end{equation}
The boundary does not move, so its equation contributes zero to $\vect{r}$. The
change in $E$ due to this vertex is then $\vect{g}\cdot d\vect{v}=\vect{g}\cdot
A^{-1}\vect{r}=\vect{y}\cdot\vect{r}$, where $\vect{y}$ is found by solving one
small linear system, $A^\trans\vect{y}=\vect{g}$. Using the expression for
$r_j$, the contributions of this vertex to the derivatives of $E$ with respect
to the generators are
\begin{equation}
  2y_j(\vect{p}_j-\vect{v})
  \;\text{ for } \frac{\partial E}{\partial \vect{p}_j}
  \;(j\ge2),
  \qquad
  -2\Bigl(\sum_{j\ge2} y_j\Bigr)(\vect{p}_1-\vect{v})
  \;\text{ for } \frac{\partial E}{\partial \vect{p}_1}.
  \label{eq:vertex-gradient}
\end{equation}
The sum is over the generators only, since an equation for the boundary
contributes nothing to $\vect{r}$. The contributions of all of the vertices are
added together. If a signature has more than $d+1$ entries, the system
$A\vect{v}=\vect{b}$ has more equations than unknowns. Since all of the
generators in the signature are equidistant from the vertex to within rounding
error, the equations are consistent, and $\vect{y}$ is taken as the
least-squares solution of $A^\trans\vect{y}=\vect{g}$ with the smallest norm.

These calculations are done in a single parallel loop over all of the
vertices of all of the cells, in C++. For the usual case of exactly $d+1$
entries, $\vect{y}$ is computed with explicit formulas for $2\times2$ and
$3\times3$ systems. When the tensors are on a GPU, the vertex derivatives are
copied to the CPU for this step, and the result is copied back. Together with
PyTorch's automatic differentiation of the objective function, this gives the
gradient of $E$ with respect to the generator positions. A vertex that is
shared by several cells contributes once for each of them, and these
contributions are simply added together, so the copies of a vertex never need
to be matched. In particular, if neighboring cells disagree about their shared
topology, the gradient is still correct for the geometry of each cell.

In a periodic domain, a vertex near the edge of the box may be defined by a
periodic image of a generator, rather than by the generator itself. Since all
of the generators that define a vertex are at the same distance from it, and
no generator is closer, the correct image of each generator is the one nearest
to the vertex. \vplex{} uses these images in $A$ and in
Eq.~\eqref{eq:vertex-gradient}. This requires the box to be large compared
with the cells, and if it is not, \vplex{} reports an error rather than
returning an incorrect gradient.

In a radical tessellation, Eq.~\eqref{eq:radical} shows that a vertex is at
the same power $\|\vect{v}-\vp_j\|^2-r_j^2$ with respect to each of its
generators, so the term $-(r_j^2-r_1^2)$ is added to the right-hand side of
Eq.~\eqref{eq:bisector}. This term does not depend on the generator
positions, so the matrix $A$, the vector $\vect{y}$, and the derivatives of
Eq.~\eqref{eq:vertex-gradient} are unchanged. Differentiating the new term
gives the contributions $-2y_jr_j$ to $\partial E/\partial r_j$ for $j\ge2$,
and $2\bigl(\sum_{j\ge2}y_j\bigr)r_1$ to $\partial E/\partial r_1$, so
\vplex{} can also return the gradient of $E$ with respect to the radii.

\subsection{Interfaces, analysis, and export}
\label{sec:interfaces}

\begin{figure}[!t]
  \centering
  \includegraphics[width=0.45\textwidth]{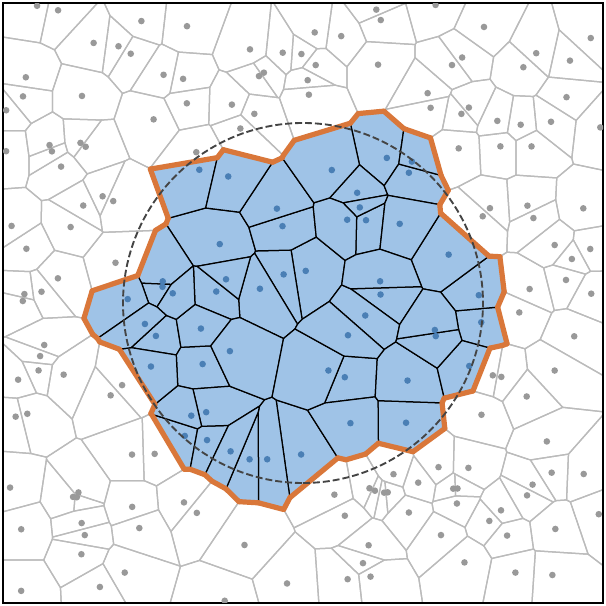}
  \captionsetup{skip=0.8em}
  \caption{The interface between inner and outer cells, extracted by \vplex.
  There are $N=200$ generators, placed at random in the unit square. The 52
  generators inside the dashed circle, of radius 0.3, are inner (blue), and
  the rest are outer (gray). The inner generators are listed first, and the
  tessellation is computed with \co{n\_inner=52}. The interface (orange) is
  the set of edges between an inner cell (shaded) and an outer cell, as
  returned by \co{surface\_mesh}. It follows the circle
  only roughly, because it is made of edges of the Voronoi cells.}
  \label{fig:interface}
\end{figure}

When the generators are divided into inner and outer sets, \vplex{} can extract
the edges (in 2D) or faces (in 3D) that separate the inner cells from the outer
cells, as shown in Fig.~\ref{fig:interface}. These form a mesh of the boundary
between the two regions, which is used in the shape meshing application of
Section~\ref{sec:sdf-shape-meshing}. Faces on the boundary of the domain can
be extracted in the same way. Further functions compute the areas, perimeters,
normals, centroids, and neighbors of the faces or cells, using PyTorch so that
these quantities can also be differentiated. Results can be exported in the VTK
\co{.vtu} format for visualization in ParaView~\cite{ahrens05}, or saved as
compressed NumPy archives (\co{.npz}) and loaded again later.

\section{Example usage}
\label{sec:examples}

Listing~\ref{lst:basic} shows the basic workflow. The function
\co{compute\_voronoi\_2d} takes a NumPy array of 2D generator positions and
returns a \co{VoronoiResult} object, whose cell data are stored in read-only
NumPy arrays. By default, the result uses the geometry representation of
Section~\ref{sec:design}, which gives the vertices, edges, and neighbors of each
cell as computed by \vpp. The \co{representation} argument selects one of the
other representations. Further functions build meshes in which each vertex is
shared between the cells that meet there, such as \co{surface\_mesh}, which
extracts the edges on the boundary of the domain. With the default
representation, these meshes use the coordinates computed by \vpp; with the
topology representation, the coordinates are recomputed from the generators, so
that they are differentiable. The 3D function \co{compute\_voronoi\_3d} works in
the same way, and the other functions work in both 2D and 3D. Each listing in
this section continues from the previous ones, using the same imports,
\co{points}, and \co{limits}.

\begin{lstlisting}[language=Python,
  caption={Basic 2D Voronoi tessellation.}, label=lst:basic]
import numpy as np
import voroplex

points = np.random.rand(1000, 2)
limits = [[0, 1], [0, 1]]

# Cell vertices, edges, and neighbors, as computed by Voro++ (the default)
result = voroplex.compute_voronoi_2d(points, limits)
print(result.areas.sum())                 # ~1.0
print(result.boundary_cells.shape)        # cells on the boundary
print(result.cell(0).vertices)            # vertices of cell 0
print(result.cell(0).neighbors)           # neighbors of cell 0

# Mesh of the edges on the boundary, with each vertex shared between cells
surface = voroplex.surface_mesh(result)
print(surface["facet_vertex_ids"].shape)  # the vertices of each edge
\end{lstlisting}

The generators can also be split into inner and outer sets, and \vplex{} can
then extract the interface between the inner and outer cells.
Listing~\ref{lst:boundary} shows this. The inner generators are stored first,
followed by the outer generators.

\begin{lstlisting}[language=Python,
  caption={Extraction of the interface between inner and outer cells.},
  label=lst:boundary]
# Cells are computed for the first 800 points, and the first 500 are inner
result = voroplex.compute_voronoi_2d(
    points, limits, n_compute=800, n_inner=500
)
interface = voroplex.surface_mesh(result)
\end{lstlisting}

Next, we use differentiation to compute a CVT, by minimizing the energy $E$ of
Eq.~\eqref{eq:cvt-uniform} with gradient-based optimization. To evaluate $E$
exactly, each cell is split into triangles, each made from the generator and one
edge of the cell, as returned by \co{cell\_fan\_simplices}. For a triangle with
corners $\vp_i$, $\vp_i+\vect{q}_1$, and $\vp_i+\vect{q}_2$, the integral of
$\|\vx-\vp_i\|^2$ over the triangle is
\begin{equation}
  \frac{a}{6}\left(\|\vect{q}_1\|^2+\|\vect{q}_2\|^2
  +\vect{q}_1\cdot\vect{q}_2\right),
  \label{eq:triangle-moment}
\end{equation}
where $a=\frac{1}{2}(q_{1,x}q_{2,y}-q_{1,y}q_{2,x})$ is the signed area of the
triangle. The function \co{cell\_fan\_simplices} orders the corners so that $a$
is positive when the generator lies inside its cell. Using the signed area,
rather than its absolute value, keeps the integrals correct even when a
generator lies outside its own cell, which can happen in the radical
tessellations described at the end of this section. Listing~\ref{lst:cvt_energy}
sums Eq.~\eqref{eq:triangle-moment} over all the triangles at once, using
PyTorch array operations rather than a Python loop over the cells.

\begin{figure}[!t]
  \centering
  \includegraphics[width=1\textwidth]{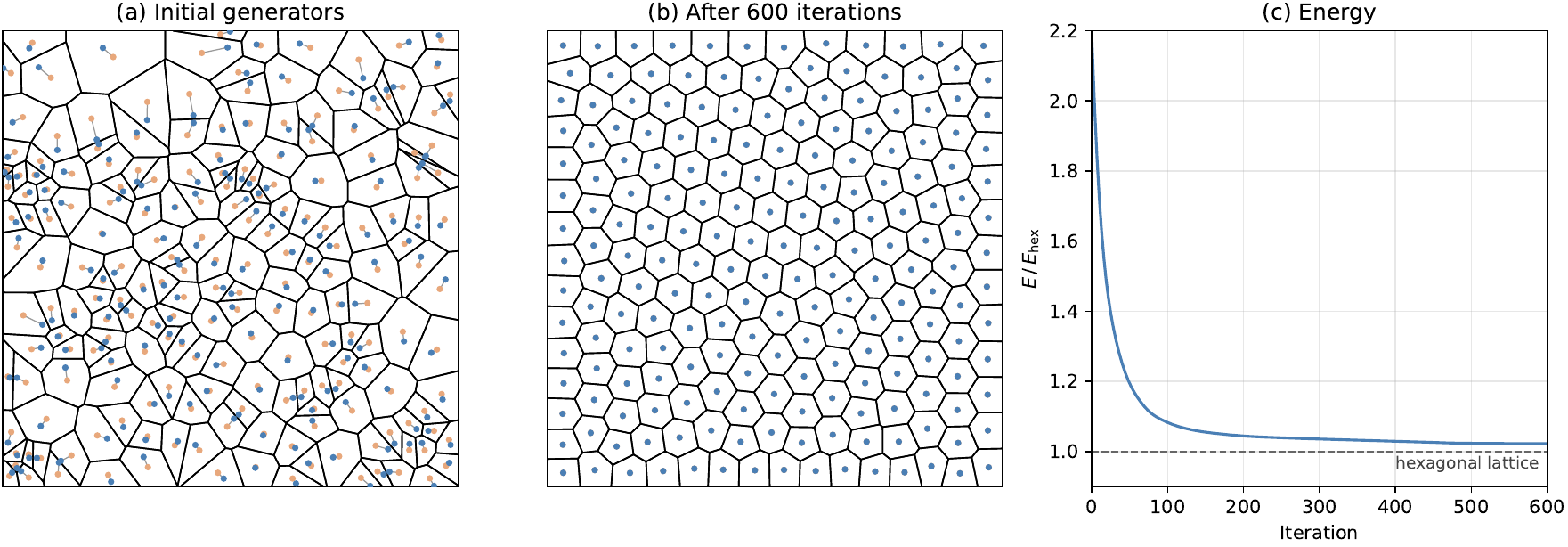}
  \captionsetup{skip=0.8em}
  \caption{Uniform-density CVT optimization in the unit square with $N=200$
    generators, computed with Listing~\ref{lst:cvt}. (a) The initial
    generators (blue), placed at random, and their Voronoi cells. Each cell's
    centroid is shown in orange, joined to its generator by a gray line.
    (b) After 600 iterations, each generator lies at the centroid of its
    cell, which is the defining property of a CVT, and most cells are close
    to regular hexagons. (c) The energy $E$ of Eq.~\eqref{eq:cvt-uniform},
    relative to $E_\text{hex}=5/(18\sqrt{3}N)$, the energy of $N$ regular
    hexagonal cells of area $1/N$. A hexagonal lattice gives the lowest
    possible energy in two dimensions~\cite{du1999centroidal}.}
	\label{fig:cvt-2d-uniform}
\end{figure}

\begin{lstlisting}[language=Python,
  caption={Uniform-density CVT energy, summed over all cells at once.},
  label=lst:cvt_energy]
import torch

def cvt_energy_2d(points, vertices, owners, simplices):
    """Fan-triangle CVT energy: one triangle per cell edge, with the
    generator as apex, summed over all cells at once."""
    generators = points[owners]                   # apex of each triangle
    q1 = vertices[simplices[:, 0]] - generators   # apex -> edge start
    q2 = vertices[simplices[:, 1]] - generators   # apex -> edge end
    # Signed triangle area and its exact second moment about the apex
    area = 0.5 * (q1[:, 0] * q2[:, 1] - q2[:, 0] * q1[:, 1])
    quadratic_term = (
        (q1 * q1).sum(dim=1) + (q2 * q2).sum(dim=1) + (q1 * q2).sum(dim=1)
    )
    return torch.sum(area * quadratic_term / 6.0)
\end{lstlisting}

Listing~\ref{lst:cvt} uses this energy for gradient-based optimization. At
each iteration, \vpp{} computes the tessellation of the current generators in
parallel, using the per-cell representation. The function
\co{cell\_vertices} returns the vertices of each cell as a PyTorch tensor,
\co{cell\_fan\_simplices} splits the cells into triangles, and
Listing~\ref{lst:cvt_energy} evaluates the energy. When
\co{loss.backward()} is called, PyTorch computes the derivatives of $E$ with
respect to the vertices, and \vplex{} converts them into derivatives with
respect to the generators, as described in Section~\ref{sec:differentiable}.
The generators then take one step using the Adam
optimizer~\cite{kingma2014adam}.

Figure~\ref{fig:cvt-2d-uniform} shows the result of Listing~\ref{lst:cvt}. The
generators start at random positions and converge to a CVT, in which each
generator lies at the centroid of its cell. The energy decreases steadily, from
2.19 to 1.02 times the energy of a hexagonal lattice. The cells next to the
boundary cannot be regular hexagons, so the hexagonal energy cannot be reached
exactly in a square. Since the density is uniform, the optimized generators are
evenly spread out, and the Voronoi cells have similar sizes and shapes. CVTs
with a density that varies in space are considered in
Section~\ref{sec:cvt-density}.

\begin{lstlisting}[language=Python,
  caption={Differentiable optimization of a 2D centroidal Voronoi
    tessellation using \vplex.},
  label=lst:cvt]
# Initial generator positions
torch.manual_seed(42)
points_t = torch.rand(200, 2, dtype=torch.float64, requires_grad=True)
limits = [[0, 1], [0, 1]]
optimizer = torch.optim.Adam([points_t], lr=1e-3)

# Differentiable CVT optimization loop
for _ in range(600):
    # Step 1: Compute the tessellation with the parallel backend
    result = voroplex.compute_voronoi_2d(
        points_t.detach().cpu().numpy(),
        limits,
        threads=8,
        representation="per_cell",
    )

    # Step 2: Get the cell vertices, differentiable with respect to points_t
    vertices = voroplex.cell_vertices(result, points_t)

    # Step 3: Split each cell into triangles
    owners, simplices = voroplex.cell_fan_simplices(result)

    # Step 4: Evaluate the exact CVT energy over all cells at once
    loss = cvt_energy_2d(points_t, vertices, owners, simplices)

    # Step 5: Backpropagate and update the generators
    optimizer.zero_grad()
    loss.backward()
    optimizer.step()

    # Keep the generators inside the computational domain
    with torch.no_grad():
        points_t.clamp_(1e-4, 1.0 - 1e-4)
\end{lstlisting}

\begin{figure}[!t]
  \centering
  \includegraphics[width=1\textwidth]{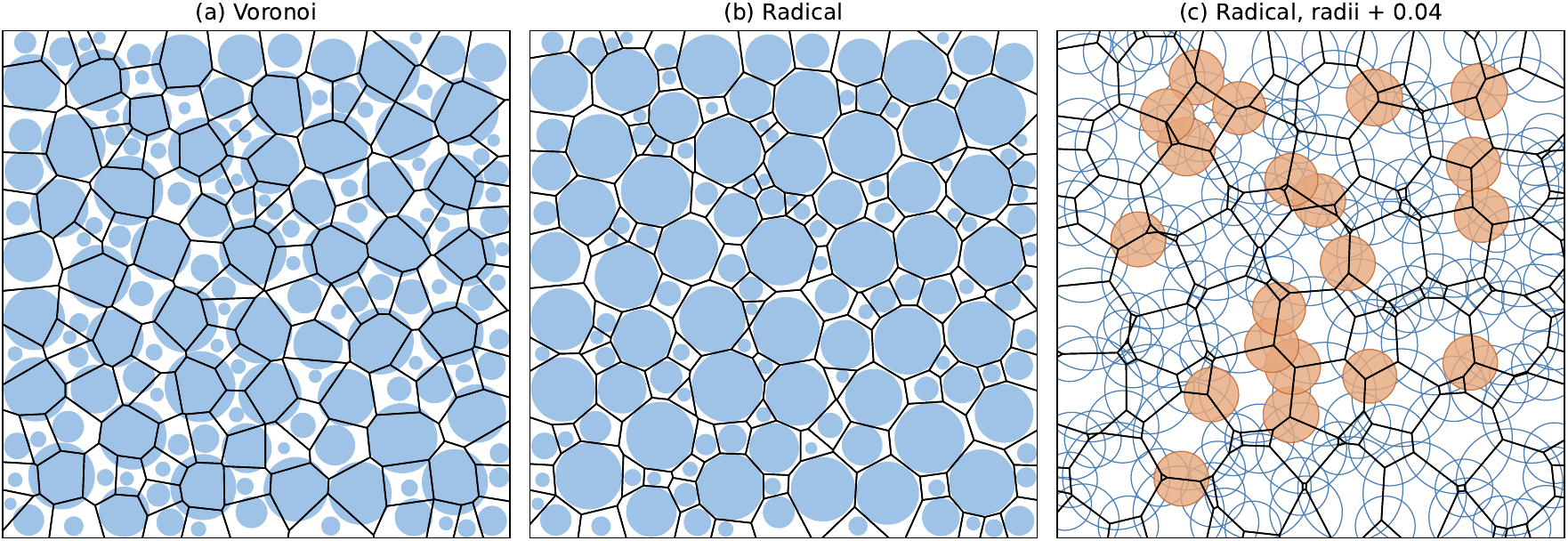}
  \captionsetup{skip=0.8em}
  \caption{Tessellations of 153 non-overlapping disks with radii between
    0.012 and 0.07 in the unit square. (a) The standard Voronoi tessellation of
    the disk centers. (b) The radical tessellation. (c) The radical tessellation
    with every radius increased by $t=0.04$, with the enlarged disks shown as
    circles. The 20 disks with empty cells are shaded.}
  \label{fig:radical}
\end{figure}

Finally, Listing~\ref{lst:radical} computes a radical tessellation, by giving a
radius for each generator. Figure~\ref{fig:radical} shows the difference for a
packing of 153 disks whose radii vary by a factor of six. The faces of the
standard Voronoi tessellation cut through many of the larger disks
(Fig.~\ref{fig:radical}(a)), while the faces of the radical tessellation lie in
the gaps between the disks (Fig.~\ref{fig:radical}(b)). Increasing all of the
radii by the same amount $t$ changes $-r_i^2$ in Eq.~\eqref{eq:radical} to
$-r_i^2-2tr_i-t^2$. The last term is the same for every disk, and the middle
term is most negative for the largest disks, so the change favors them. Their
cells grow, and some of the smaller disks, which now lie inside the larger ones,
have empty cells (Fig.~\ref{fig:radical}(c)). As $t$ increases further, the cell
of the largest disk eventually fills the whole domain. For differentiation, the
radii can also be passed to \co{cell\_vertices} as a PyTorch tensor, and
gradients then flow to them as well as to the positions.

\begin{lstlisting}[language=Python,
  caption={Radical tessellation of particles with different radii.},
  label=lst:radical]
radii = 0.03 * np.random.rand(1000)    # one radius for each generator
result = voroplex.compute_voronoi_2d(points, limits, radii=radii)
print(result.areas.sum())              # ~1.0
print(np.sum(result.areas == 0))       # number of empty cells
\end{lstlisting}

\section{Performance evaluation}
\label{sec:performance}

The CPU benchmarks were run on two identical computers, each with two Intel
Xeon E5-2650L v4 processors and 256\,GiB of memory. Each processor has 14
cores, giving 28 cores in total. The base clock frequency of 1.7\,GHz is the
frequency that the processors can always sustain. Turbo Boost raises it when
possible: to 2.5\,GHz when one or two cores are busy, and to 2.0\,GHz when six
or more cores are busy~\cite{lu2022}. Turbo Boost was left on, as it is on most
computers, except where stated below. The GPU benchmarks were run on a third
computer with the same processors and an NVIDIA GeForce RTX 2080 Ti GPU with
11\,GiB of memory. All three computers run Ubuntu 26.04 with GCC 15.2, Python
3.14.4, NumPy 2.5.4, SciPy 1.18.1, and PyTorch 2.14.1 (with CUDA 12.6 for the
GPU). In every test, the generators were drawn uniformly at random from the
unit square or the unit cube. Each measurement was made in a new process and
repeated five times with different generators, and the median is reported.
Each thread was pinned to its own physical core, with threads 1--14 on the
first processor and threads 15--28 on the second. The average clock frequency
during each measurement was found by counting clock cycles with the
processor's hardware counters, using the Linux \co{perf} tool.

\subsection{Comparison with existing Python Voronoi libraries}

\begin{table}[h]
\centering
\small
\begin{tabular}{llrrrr}
\toprule
 & & \multicolumn{2}{c}{$N=10^5$} & \multicolumn{2}{c}{$N=10^6$} \\
\cmidrule(lr){3-4}\cmidrule(lr){5-6}
Library & Threads & 2D & 3D & 2D & 3D \\
\midrule
SciPy & 1 & 1.02 & 5.05 & 13.53 & 57.17 \\
pyvoronoi & 1 & 3.25 & -- & 46.60 & -- \\
pyvoro & 1 & 19.34 & 20.72 & 236.5 & 268.2 \\
\midrule
\multirow{2}{*}{\vplex{} (geometry)} & 1 & 0.229 & 2.53 & 2.62 & 26.01 \\
 & 28 & 0.027 & 0.132 & 0.295 & 1.47 \\
\cmidrule{2-6}
\multirow{2}{*}{\vplex{} (geometry, no neighbors)}
  & 1 & 0.214 & 2.38 & 2.49 & 24.57 \\
 & 28 & 0.026 & 0.127 & 0.288 & 1.41 \\
\cmidrule{2-6}
\multirow{2}{*}{\vplex{} (per-cell)} & 1 & 0.281 & 2.77 & 3.18 & 28.56 \\
 & 28 & 0.031 & 0.146 & 0.346 & 1.65 \\
\cmidrule{2-6}
\multirow{2}{*}{\vplex{} (topology)} & 1 & 0.261 & 2.68 & 2.95 & 27.54 \\
 & 28 & 0.029 & 0.139 & 0.320 & 1.52 \\
\bottomrule
\end{tabular}
\caption{Time to compute the Voronoi tessellation of $N$ generators, in seconds.
pyvoronoi has no 3D version. The four \vplex{} rows are the geometry
representation with and without neighbor information, and the per-cell and
topology representations. Each value is the median of five independent trials,
each with different random generators.}
\label{tab:comparison}
\end{table}

\begin{table}[h]
\centering
\small
\begin{tabular}{rrrrr}
\toprule
 & \multicolumn{2}{c}{2D} & \multicolumn{2}{c}{3D} \\
\cmidrule(lr){2-3}\cmidrule(lr){4-5}
$N$ & Time (s) & Memory (GiB) & Time (s) & Memory (GiB) \\
\midrule
$10^{5}$ & 19.34 & 21.1 & 20.72 & 21.8 \\
$2\times10^{5}$ & 38.76 & 42.1 & 41.66 & 43.6 \\
$5\times10^{5}$ & 96.70 & 106.8 & 105.1 & 108.9 \\
$10^{6}$ & 236.5 & 213.5 & 268.2 & 217.7 \\
\bottomrule
\end{tabular}
\caption{Time and peak memory used by pyvoro to compute the Voronoi tessellation
of $N$ generators. Each value is the median of five independent trials, each
with different random generators.}
\label{tab:pyvoro}
\end{table}

Table~\ref{tab:comparison} compares \vplex{} with three Python libraries for
Voronoi tessellations, for $N=10^5$ and $N=10^6$ generators.
\co{scipy.spatial.Voronoi}~\cite{virtanen2020scipy} is the most widely used. It
computes the tessellation from a convex hull in one dimension higher, using
Qhull~\cite{qhull_lib}. pyvoronoi~\cite{ancelin2025pyvoronoi} uses the
sweep-line algorithm~\cite{fortune86} of the C++ library
Boost.Polygon~\cite{simonson2009geometry,sydorchuk_boost_voronoi}, and works
only in 2D. pyvoro~\cite{pyvoro} is an interface to the original,
single-threaded \vpp, which returns each cell as a Python dictionary of lists.
All three libraries use one thread. SciPy and pyvoronoi compute the unbounded
Voronoi diagram, in which the cells at the edge extend to infinity, while
\vplex{} and pyvoro clip every cell to a box. For pyvoronoi, the time includes
adding the points, building the diagram, and copying the cells into Python;
building the diagram alone takes about a third of the time.

On one thread, \vplex{} with the geometry representation is 4.4--5.2 times
faster than SciPy in 2D and 2.0--2.2 times faster in 3D, and 14--18 times
faster than pyvoronoi in 2D. On 28 threads, it is 38--46 times faster than
SciPy in 2D and 38--39 times faster in 3D. pyvoro is much slower: for
$N=10^6$, it is 90 times slower than \vplex{} on one thread in 2D, and 10 times
slower in 3D. Table~\ref{tab:pyvoro} shows that pyvoro also uses about 230\,KiB
of memory per cell, so that it needs 214\,GiB in 2D and 218\,GiB in 3D for
$N=10^6$. This is because pyvoro computes all of the cells, and stores a copy
of each \vpp{} cell object, including its working memory, before it converts
any of them into Python objects.

\vpp{} has two classes for computing a cell: one that records the neighbor of
each face, and a simpler one that does not. By default, \vplex{} records the
neighbors, which are needed for the neighbor IDs, the boundary flags, and the
per-cell and topology representations. Users who only need the geometry of
each cell, such as its volume, can pass \co{neighbors=False} to use the simpler
class, which is 2--7\% faster. The per-cell representation takes 9--23\%
longer than the geometry representation, because it also computes and stores
the signature of every vertex, and the topology representation takes 3--14\%
longer. Since most users do not need derivatives, the geometry representation
with neighbors is the default.

\subsection{Parallel performance}

\begin{figure}[t]
\centering
\includegraphics[width=\textwidth]{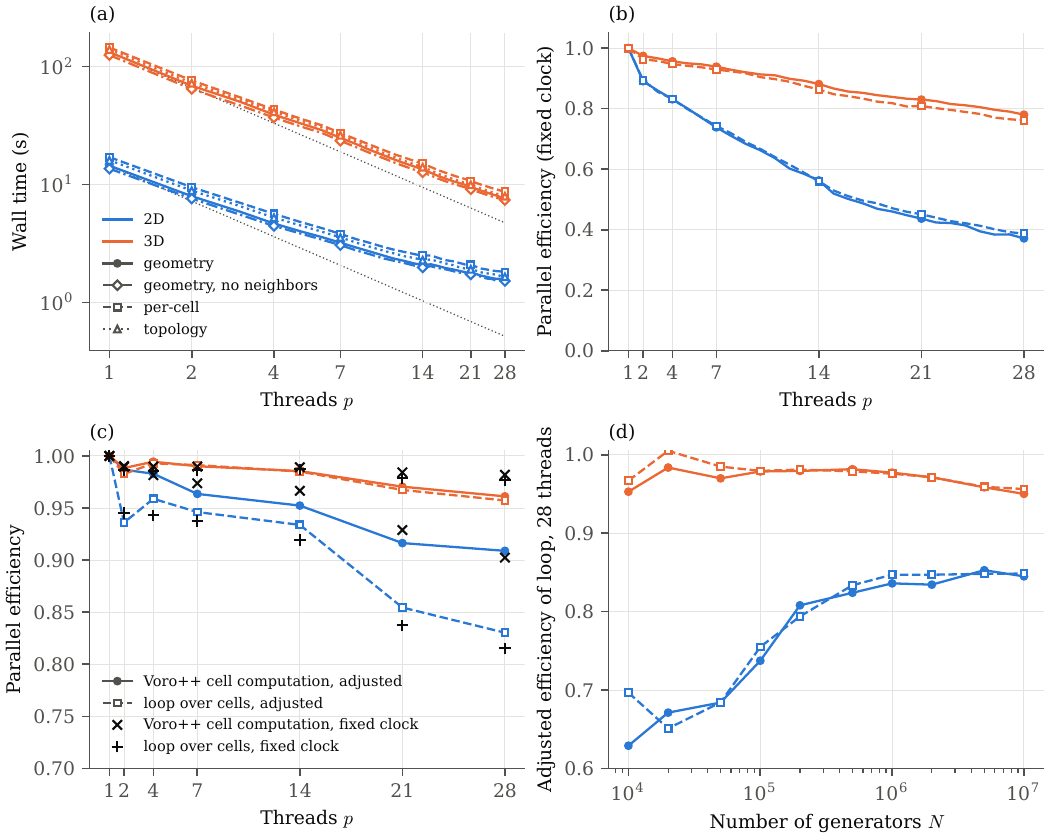}
\caption{Parallel performance of \vplex{} in 2D and 3D. (a)~Wall time with Turbo
Boost on, against the number of threads $p$, for $N=5\times10^6$ generators and
the four ways of computing the tessellation. The dotted lines show perfect
scaling. (b)~Parallel efficiency $T_1/(pT_p)$ of a whole call at a fixed clock
frequency (Turbo Boost off), where $T_p$ is the wall time on $p$ threads, for
the geometry and per-cell representations. (c)~Parallel efficiency of the loop
over the cells, and of the \vpp{} cell computation within it, for the geometry
representation: adjusted for the clock frequency with Turbo Boost on (lines,
Eq.~\eqref{eq:adjusted}), and measured at a fixed clock frequency (crosses).
(d)~Adjusted parallel efficiency of the loop over the cells on 28 threads,
against the number of generators $N$. Each value is the median of five
independent trials, each with different random generators.}
\label{fig:parallel}
\end{figure}

\begin{figure}[t]
\centering
\includegraphics[width=\textwidth]{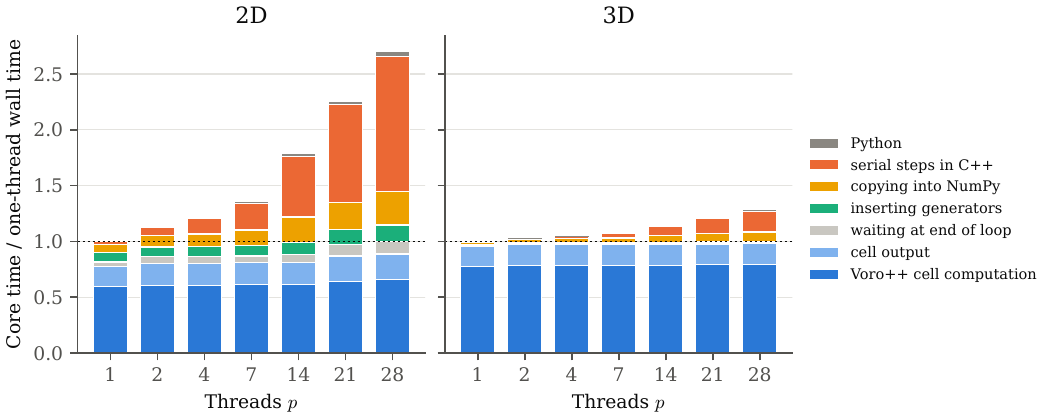}
\caption{Where the time goes in a call of \vplex{}, for the geometry
representation with $N=5\times10^6$ generators, at a fixed clock frequency
(Turbo Boost off). Each bar is the core time of one part of the calculation,
which is its wall time multiplied by the number of threads $p$, divided by the
wall time of the whole call on one thread. The parts are stacked, so that the
total height of the bars for each $p$ is $1/E$, where $E$ is the parallel
efficiency, and the dotted line at 1 shows perfect scaling. A part that scales
perfectly keeps the same height as $p$ increases. Each value is the median of
five independent trials, each with different random generators.}
\label{fig:breakdown}
\end{figure}

Turbo Boost makes parallel efficiency harder to measure, because one thread
runs at 2.5\,GHz, while 28 threads run at 2.0\,GHz. Earlier
work~\cite{lu2022} on the multi-threaded \vpp{} handled this with the adjusted
parallel efficiency
\begin{equation}
  E^\text{adj}(p) = \frac{T_1}{p\,T_p}\,\frac{f_1}{f_p},
  \label{eq:adjusted}
\end{equation}
where $f_p$ is the average clock frequency of the busy threads on $p$ threads,
measured from the number of clock cycles. This adjustment is accurate only if
all of the threads are busy for the whole of the time measured~\cite{lu2022}.
That holds for the loop over the cells, but not for a whole call of \vplex{},
which also has steps that run on one thread. During these steps, the one busy
core runs at up to 2.5\,GHz while the others are idle, so that dividing by the
average frequency $f_p$ overestimates the efficiency. In 2D, for example,
Eq.~\eqref{eq:adjusted} gives an efficiency of 0.44 for a whole call on 28
threads, but the measured efficiency at a fixed clock frequency is 0.37. We
therefore follow the approach of Ref.~\cite{lu2022} for calculations that mix
serial and parallel steps: we switched Turbo Boost off on the GPU computer,
which has the same processors, so that the clock frequency stays at 1.7\,GHz
whatever the number of threads, and measured the parallel efficiency
there. We report the wall times with Turbo Boost on, since these are what a
user would see.

Figure~\ref{fig:parallel}(a,b) shows how the time to compute the tessellation
of $N=5\times10^6$ generators depends on the number of threads. With Turbo
Boost on, 28 threads are 17.2 times faster than one thread in 3D, and 9.2 times
faster in 2D. At a fixed clock frequency, the parallel efficiency of a whole
call on 28 threads is 0.78 in 3D and 0.37 in 2D. The per-cell representation
behaves in almost the same way, with efficiencies of 0.76 and 0.39.

Figure~\ref{fig:parallel}(c) shows that most of the loss of efficiency is not in
the parallel computation of the cells. The \vpp{} cell computation has an
adjusted efficiency of 0.96 on 28 threads in 3D, and 0.91 in 2D, and the
measurements at a fixed clock frequency agree (0.98 and 0.90). This is
consistent with the results for the multi-threaded \vpp{}~\cite{lu2022}. The
whole loop over the cells, which also copies each cell's vertices and faces into
\vplex's buffers, has an efficiency of 0.96 in 3D and 0.83 in 2D.

Figure~\ref{fig:breakdown} shows where the rest of the time goes. In 3D, the
\vpp{} cell computation and the output of each cell take almost the same core
time on any number of threads, as they would with perfect scaling. The core time
grows mainly in two parts: the steps that run on one thread in C++, and the copy
of the result into NumPy arrays, which runs in parallel but is only about 7
times faster on 28 threads than on one thread. The steps on one thread are
creating the \vpp{} container, which allocates the memory for each of its
roughly $10^6$ blocks, and freeing the memory of the container and of the result
at the end. On 28 threads, they take 0.9\,s in 2D and 1.2\,s in 3D, a little
more than on one thread, because freeing memory that was allocated by many
threads takes longer. Since a 2D cell is about ten times cheaper to compute than
a 3D cell, these steps take a much larger share of the time in 2D. On 28
threads, they account for 45\% of the core time in 2D, and the copy into NumPy
for another 11\%. The Python code in \vplex{} takes only 2\% of the core time,
so the limit is in these serial steps in C++, which could be made faster in the
future, for example by allocating the blocks of the container in parallel.

Figure~\ref{fig:parallel}(d) shows the adjusted efficiency of the loop over
the cells on 28 threads, for $N$ from $10^4$ to $10^7$. In 3D, it is between
0.95 and 1.0 over this whole range, so the cells are computed efficiently in
parallel even for small problems, which take only 14\,ms on 28 threads for
$N=10^4$. In 2D, the efficiency rises from 0.63 at $N=10^4$ to about 0.85 for
$N\ge10^6$. A 2D cell is cheap, so for small $N$ each thread computes only a few
hundred cells, taking about a millisecond, and the cost of starting and
coordinating the threads is probably no longer negligible. At a fixed clock
frequency, the efficiency of a whole call on 28 threads is between 0.73 and
0.85 in 3D for all $N$, and rises from 0.29 to about 0.38 in 2D.

\subsection{Differentiable CVT optimization}

\begin{table}[h]
\centering
\small
\begin{tabular}{llrrrrr}
\toprule
 & & \multicolumn{3}{c}{\vplex} & & \\
\cmidrule{3-5}
Dim. & Device & Tessellation & Forward & Backward & PyTorch & Total \\
\midrule
2D & CPU & 23 & 1.5 & 5.0 & 41 & 71 \\
2D & GPU & 25 & 4.8 & 7.6 & 8.2 & 46 \\
3D & CPU & 124 & 20 & 33 & 1248 & 1425 \\
3D & GPU & 123 & 38 & 88 & 58 & 306 \\
\bottomrule
\end{tabular}
\caption{Time per iteration, in milliseconds, of the CVT optimization with
$N=10^5$ generators. For each of five independent trials, each with different
random generators, the median over the 500 iterations was found, and the median
of these five values is reported. The tessellation uses 28 threads. The \vplex{}
columns are the tessellation, the forward pass (\co{cell\_vertices} and
\co{cell\_fan\_simplices}), and the vertex derivative in the backward pass
(Section~\ref{sec:differentiable}). The PyTorch column is the rest: the energy,
its derivatives with respect to the vertices, and the optimizer step.}
\label{tab:cvt}
\end{table}

As an example of an optimization, we timed the CVT calculation of
Section~\ref{sec:examples}, using the energy of
Eq.~\eqref{eq:cvt-uniform}, for $N=10^5$ generators over 500 iterations of the
Adam optimizer, in 2D and 3D. In each iteration, \vplex{} computes the
tessellation on the CPU with 28 threads, and the rest of the calculation runs
either on the CPU, also with 28 threads, or on the GPU. Table~\ref{tab:cvt}
shows the time taken by each part of an iteration.

On the CPU, most of the time is spent in PyTorch, evaluating the energy and
its derivatives with respect to the vertices: 1.25\,s of the 1.43\,s per
iteration in 3D. The parts done by \vplex{} take only 12\% of the time in 3D,
and the vertex derivative in the backward pass takes 33\,ms. On the GPU, the
PyTorch part is about 22 times faster in 3D, and an iteration takes 0.31\,s.
The tessellation then takes the largest share of the time: 40\% in 3D and 55\%
in 2D. The vertex derivative takes longer on the GPU than on the CPU (88\,ms
instead of 33\,ms), because it is computed on the CPU, and the data must be
copied from the GPU and back.

PyTorch also limits how well the CPU calculation uses many threads. In a
separate measurement in 3D, with one trial of 50 iterations, going from one
thread to 28 threads makes the tessellation 18 times faster and the vertex
derivative 13 times faster, but the PyTorch part of the backward pass only 4.0
times faster. As a result, a whole iteration becomes only 6.6 times faster,
taking 9.9\,s on one thread and 1.5\,s on 28 threads.

The size of the problem that fits on the GPU is limited by its memory. The
peak memory allocated by PyTorch grows in proportion to $N$. In 3D,
$N=5\times10^5$ generators used 6.9\,GiB. $N=7\times10^5$ would need about
9.7\,GiB, and it did not fit in the GPU's 11\,GiB, since part of this memory
is also needed by the CUDA runtime and PyTorch's memory allocator. In 2D,
$N=2\times10^6$ used 2.4\,GiB. On the CPU, which has much more memory, an
iteration with $N=10^6$ in 3D takes 14\,s.

\FloatBarrier

\section{Differentiable Voronoi optimization applications}
\label{sec:applications}

This section presents three applications of differentiable Voronoi
tessellations. In each one, an objective function is computed from the Voronoi
cells using PyTorch, and its gradient with respect to the generator positions is
used to optimize the generators.

\subsection{Density-weighted CVT}
\label{sec:cvt-density}

Section~\ref{sec:examples} computed a CVT for a uniform density, using the
energy of Eq.~\eqref{eq:cvt-uniform}. With a density field $\rho(\vx)\ge0$,
the energy becomes
\begin{equation}
  E_{\rho}
  =
  \sum_{i=1}^{N}
  \int_{V_i}
  \rho(\vx)
  \|\vx-\vp_i\|^2
  \,d\vx.
  \label{eq:cvt-density}
\end{equation}
At a minimum of this energy, each generator lies at the density-weighted
centroid of its cell,
\begin{equation}
  \vc_i
  =
  \frac{
    \int_{V_i}
    \rho(\vx)\vx
    \,d\vx
  }{
    \int_{V_i}
    \rho(\vx)
    \,d\vx
  }.
  \label{eq:weighted-centroid}
\end{equation}
More generators are therefore placed where $\rho$ is large, and the cells there
are smaller. Here $\rho$ is the weight in the CVT energy, rather than the number
density of the generators. In the asymptotic regime, the local cell length scale
is proportional to $\rho^{-1/(d+2)}$, and the generator number density is
proportional to $\rho^{d/(d+2)}$, where $d$ is the spatial
dimension~\cite{du1999centroidal}. Therefore, a target relative spacing $h(\vx)$
is represented by the energy density $\rho(\vx)=h(\vx)^{-(d+2)}$. Unlike the
energy for a uniform density, the density-weighted energy cannot in general be
computed exactly, so it is approximated by numerical quadrature. In 2D, each
cell is split into a fan of triangles, and a three-point quadrature rule is used
on each triangle. In 3D, each face of a cell is split into triangles that share
the centroid of its vertices, and each triangle is joined to the generator to
make a tetrahedron, on which a four-point quadrature rule is used. The
tetrahedra do not depend on how the vertices of each face are numbered, which
can differ between runs with several threads, so the result is reproducible.
These calculations use PyTorch operations, so that the gradient of the energy
with respect to the generator positions can be computed, and the generators are
optimized with Adam~\cite{kingma2014adam}.

\subsubsection{Two-dimensional examples}

We present three examples in the unit square. The first specifies a target
relative spacing through the sizing field
\begin{equation}
    h(\vx)=1+4x+4y.
    \label{eq:sizing-2d-linear-strong}
\end{equation}
Using the spacing--density relation above with $d=2$, we define the
corresponding energy density as
\begin{equation}
    \rho(\vx)=h(\vx)^{-4}
    =(1+4x+4y)^{-4}.
    \label{eq:density-sizing-relationship-2d}
\end{equation}
The target relative spacing increases from 1 at the origin to 9 at the opposite
corner. The cells are therefore expected to be smallest near the origin and to
become larger as $x+y$ increases. The prescribed spacing describes an asymptotic
relative length scale; it does not impose an exact diameter on each individual
cell. The other two examples specify the energy density $\rho(\vx)$ directly,
without a sizing-field conversion.

The second example uses a density with two Gaussian peaks,
\begin{equation}
    \rho(\vx) ={} 0.12
    +4.8\exp\left(
        -\frac{\lVert\vx-\vect{a}_1\rVert^2}{2\sigma_1^2}
    \right)
    +3.6\exp\left(
        -\frac{\lVert\vx-\vect{a}_2\rVert^2}{2\sigma_2^2}
    \right),
\label{eq:density-2d-gauss}
\end{equation}
where
\begin{equation}
    \vect{a}_1=(0.30,0.72),
    \qquad
    \vect{a}_2=(0.74,0.32),
    \qquad
    \sigma_1=0.10,
    \qquad
    \sigma_2=0.13.
\end{equation}
This tests two separate regions of high density.

The third example concentrates the density in a narrow band around a
five-petal curve,
\begin{equation}
    \rho(\vx)=0.10
    +3.0\exp\left(
        -\frac{\left(r-r_{\ast}(\theta)\right)^2}{2\sigma_r^2}
    \right),
    \label{eq:density-2d-flower}
\end{equation}
where
\begin{equation}
    r_{\ast}(\theta)=0.30+0.075\cos(5\theta+0.35),
\end{equation}
$(r,\theta)$ are polar coordinates centered at $(0.5,0.5)$, and
$\sigma_r=0.045$. This tests whether the generators can follow a thin, curved
region of high density.

\begin{figure}[!htbp]
    \centering
    \includegraphics[width=\linewidth]
    {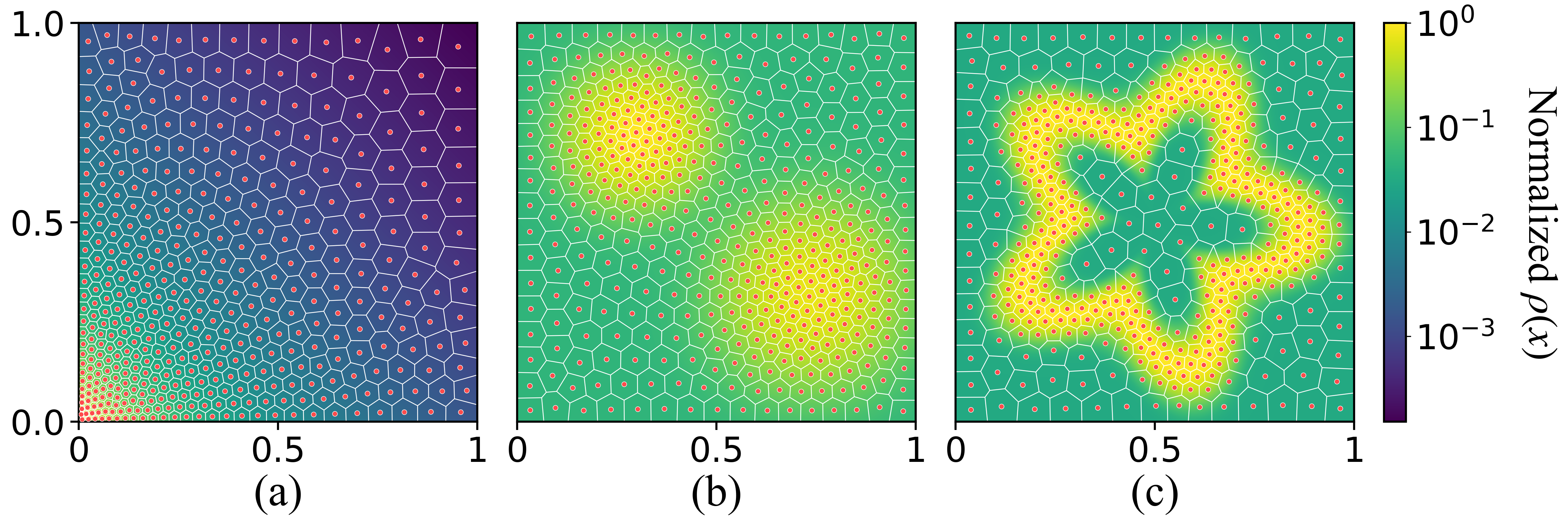}

\caption{ Two-dimensional density-weighted CVTs with $N=400$ generators.
(a)~$\rho(\vx)=h(\vx)^{-4}$ for the linear sizing field $h$ of
Eq.~\eqref{eq:sizing-2d-linear-strong}. (b)~Two Gaussian peaks
(Eq.~\eqref{eq:density-2d-gauss}). (c)~A band around a five-petal curve
(Eq.~\eqref{eq:density-2d-flower}). Red points denote the optimized generators,
and white lines denote the Voronoi cell boundaries. In each panel, the
background density is normalized by its own maximum,
$\widetilde{\rho}=\rho/\rho_{\max}$, and displayed using a shared logarithmic
color scale.}

\label{fig:cvt-2d}

\end{figure}

All three examples start from the same $N=400$ random generator positions,
sampled independently and uniformly from $[0.05,0.95]^2$ using random seed 42.
The generators are optimized using Adam with a learning rate of $0.005$ for 400
iterations. Figure~\ref{fig:cvt-2d} shows the optimized tessellations. With the
linear sizing field, the cells become larger away from the origin. With the two
Gaussian peaks, the generators gather in the two regions of high density, while
some remain in the background. With the five-petal density, the generators
follow the narrow band around the curve.

\FloatBarrier
\subsubsection{Three-dimensional examples}

We consider three examples in the unit cube, all using the same $N=1500$
initial generators, sampled independently and uniformly from $[0.05,0.95]^3$
with random seed 42. The first uses a uniform density, $\rho(\vx)=1$, for which
the cells should all have similar sizes. The second uses a sizing field
$h(\vx)$, which, as in 2D, is the desired relative length scale of the cells, so
that the volume of a cell is proportional to $h^3$. It is converted into a
density by
\begin{equation}
\rho(\vx)=h(\vx)^{-5},
\label{eq:sizing-to-density-3d}
\end{equation}
with
\begin{equation}
h(\vx)
=
1+x+y+z,
\label{eq:sizing-3d-linear}
\end{equation}
so that the cells should be smallest at the origin and grow linearly along the
diagonal of the cube.

The third example makes the cells smaller near the edges of the cube. Let
\begin{equation}
d_{\mathcal{E}}(\vx)
=
\min_{e\in\mathcal{E}}
\operatorname{dist}(\vx,e)
\label{eq:distance-to-cube-edges}
\end{equation}
be the distance from $\vx$ to the nearest of the twelve edges $\mathcal{E}$ of
the cube. Its largest value is $d_{\max}=\sqrt{2}/2$, at the center of the
cube. The sizing field is
\begin{equation}
h(\vx)
=
h_{\min}
+
\left(h_{\max}-h_{\min}\right)\frac{d_{\mathcal{E}}(\vx)}{d_{\max}},
\label{eq:sizing-3d-edge}
\end{equation}
with $h_{\min}=0.6$ and $h_{\max}=1.2$, so that it is smallest on the edges
and grows linearly with the distance from them.

\begin{figure}[htbp]
    \centering

    \includegraphics[width=\linewidth]
    {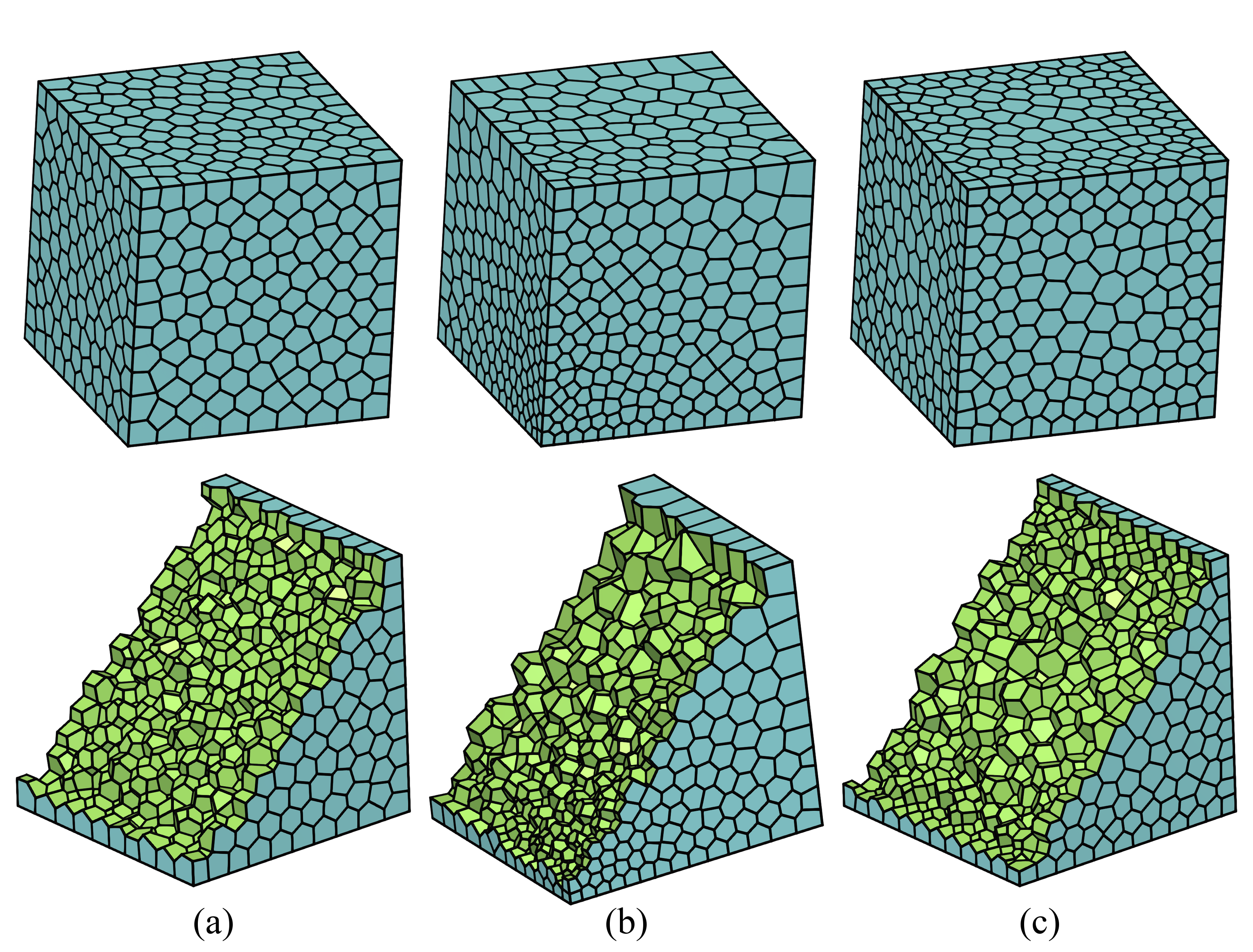} \caption{Three-dimensional
density-weighted CVTs with $N=1500$ generators. (a)~Uniform density.
(b)~Linear sizing field (Eq.~\eqref{eq:sizing-3d-linear}). (c)~Sizing field
that is smallest on the edges of the cube (Eq.~\eqref{eq:sizing-3d-edge}).
The top row shows the whole tessellation. In the bottom row, the whole cells on
one side of a diagonal plane are removed, so that the cells inside can be
seen.}
    \label{fig:cvt-3d}
\end{figure}

The generators are optimized with Adam, with a learning rate that decreases on
a cosine schedule, from $0.002$ to $0.0002$ for the first and third examples,
and from $0.0008$ to $0.00005$ for the second. The optimization stops when the
range of the energy over the last 30 iterations, divided by its mean, falls
below $10^{-4}$, or after at most 1600 iterations. This took 560, 1317, and 478
iterations for the three examples.

Figure~\ref{fig:cvt-3d} shows the optimized tessellations. With a uniform
density, the cells have similar sizes throughout the cube. With the linear
sizing field, the cell size varies smoothly along the diagonal, and with the
third sizing field, the generators gather along the twelve edges of the cube.

\FloatBarrier

\subsection{2D Voronoi shape meshing from signed distance functions}
\label{sec:sdf-shape-meshing}

We next use \vplex{} to mesh a 2D shape that is given by a signed distance
function (SDF)~\cite{persson2004simple}. Let $\phi$ be the SDF of the shape,
which is negative inside the shape and positive outside. It divides the domain
$\Omega$ into an interior region, an exterior region, and the boundary
$\Gamma$ of the shape,
\begin{equation}
\begin{aligned}
  \Omega_{\mathrm{in}}
  &=
  \{\vx\in\Omega:\phi(\vx)\leq 0\},\\
  \Omega_{\mathrm{out}}
  &=
  \{\vx\in\Omega:\phi(\vx)>0\},\\
  \Gamma
  &=
  \{\vx\in\Omega:\phi(\vx)=0\}.
\end{aligned}
\label{eq:sdf-domain-partition}
\end{equation}
The aim is to place the generators so that the edges between the cells inside
and outside the shape follow $\Gamma$.

\begin{figure}[H]
    \centering
    \includegraphics[width=0.6\linewidth]
    {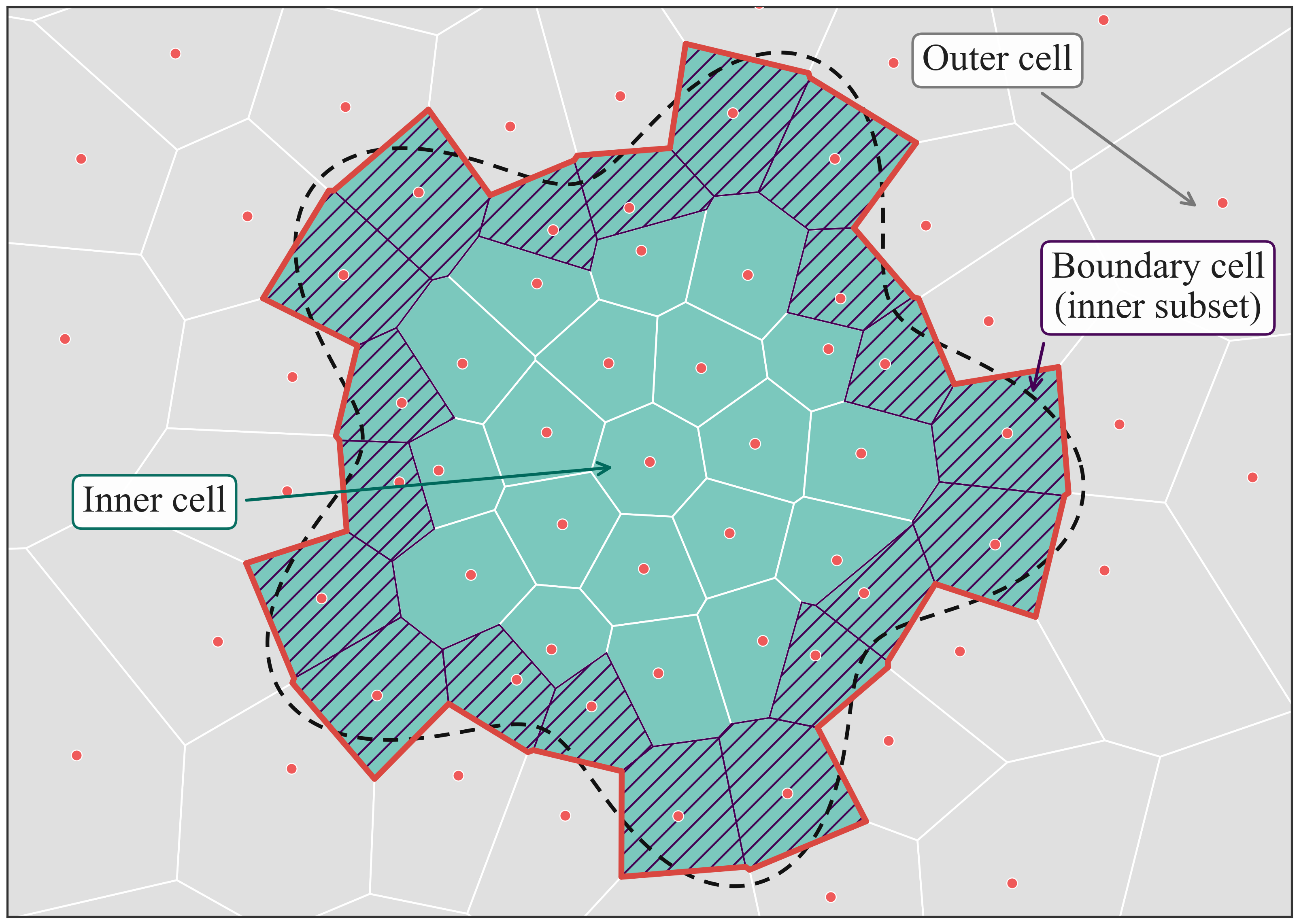}
    \caption{The classification of cells used to mesh a shape given by an SDF.
    Teal cells are inner cells, and the boundary cells, which are inner cells
    next to the interface, are marked with purple hatching. Gray cells are
    outer cells. Coral points are the generators, solid coral segments are the
    interface edges $\mathcal{E}_{\Gamma}$, and the black dashed curve is the
    boundary $\Gamma$ of the shape.}
    \label{fig:sdf-cell-classification}
\end{figure}

Each generator is labeled as inner if $\phi(\vp_i)\leq 0$, and as outer
otherwise, and the generators are reordered so that the $n_{\mathrm{in}}$
inner generators come first. When $n_{\mathrm{in}}$ is passed to \vplex{}
(Section~\ref{sec:interfaces}), it finds the set of edges between the inner and
outer cells,
\begin{equation}
  \mathcal{E}_{\Gamma}
  =
  \left\{\text{edge }
  e:
  \operatorname{owner}(e)\leq n_{\mathrm{in}},
  \quad
  \operatorname{neighbor}(e)> n_{\mathrm{in}}
  \right\},
  \label{eq:sdf-interface-edges}
\end{equation}
where $\operatorname{owner}(e)$ and $\operatorname{neighbor}(e)$ are the indices
of the generators of the two cells that share the edge $e$. The boundary cells
are the inner cells that have an edge in $\mathcal{E}_{\Gamma}$,
\begin{equation*}
  \mathcal{B}_{\Gamma}
  =
  \left\{
    \operatorname{owner}(e):
    e\in\mathcal{E}_{\Gamma}
  \right\}.
\end{equation*}
The outer cells next to the interface are not boundary cells. These
definitions are shown in Fig.~\ref{fig:sdf-cell-classification}.

At each iteration, \vplex{} recomputes the tessellation and the interface edges,
using the per-cell representation, so that the vertices of these edges are
differentiable functions of the generator positions. For an interface edge
$e\in\mathcal{E}_{\Gamma}$ with endpoints $\vect{v}_{e,0}$ and $\vect{v}_{e,1}$,
let
\begin{equation}
  \vect{m}_e
  =
  \frac{1}{2}
  \left(
    \vect{v}_{e,0}
    +
    \vect{v}_{e,1}
  \right)
  \label{eq:sdf-edge-midpoint}
\end{equation}
be its midpoint. The objective function is
\begin{equation}
\begin{aligned}
  \mathcal{L}_{\Gamma}
  &=
  \lambda_{\mathrm{mid}}\mathcal{L}_{\mathrm{mid}}
  +
  \lambda_{\mathrm{end}}\mathcal{L}_{\mathrm{end}}
  +
  \lambda_{\mathrm{in}}\mathcal{L}_{\mathrm{in}}
  +
  \lambda_{\mathrm{area}}\mathcal{L}_{\mathrm{area}}
  +
  \lambda_{\mathrm{circ}}\mathcal{L}_{\mathrm{circ}},
\end{aligned}
\label{eq:sdf-boundary-objective}
\end{equation}
where the weights $\lambda$ are constants. The five terms are described below.

\textbf{Matching the shape.} The first two terms move the interface edges onto
$\Gamma$, by penalizing the SDF at the midpoint and at the two endpoints of each
edge,
\begin{equation}
\begin{aligned}
  \mathcal{L}_{\mathrm{mid}}
  &=
  \sum_{e\in\mathcal{E}_{\Gamma}}
  \phi(\vect{m}_e)^2,\\
  \mathcal{L}_{\mathrm{end}}
  &=
  \sum_{e\in\mathcal{E}_{\Gamma}}
  \left(
    \phi(\vect{v}_{e,0})^2
    +
    \phi(\vect{v}_{e,1})^2
  \right).
\end{aligned}
\label{eq:sdf-mid-end-loss}
\end{equation}

\textbf{The inner cells.} The third term is the density-weighted CVT energy of
the inner cells,
\begin{equation}
  \mathcal{L}_{\mathrm{in}}
  =
  \sum_{i=1}^{n_{\mathrm{in}}}
  \int_{V_i}
  \rho_{\Gamma}(\vx)
  \|\vx-\vp_i\|^2
  \,d\vx,
  \label{eq:sdf-inner-cvt-loss}
\end{equation}
which keeps the inner cells regular. As in Section~\ref{sec:cvt-density}, it
is evaluated with a three-point quadrature rule on the triangles of each cell.
The density $\rho_{\Gamma}$ is chosen so
that the cells are small near $\Gamma$ and larger away from it. It is defined
from a desired cell length scale $\mu_{\Gamma}(\vx)$, which changes from
$\mu_{\min}$ on $\Gamma$ to $\mu_{\max}$ at a distance $w_{\Gamma}$ or more from
it. Since $\phi$ is a signed distance function, $|\phi(\vx)|$ is the distance
from $\vx$ to $\Gamma$. (For the flower and the peanut below, $\phi$ is only
approximately a distance, which is enough here.) We therefore define
\begin{equation}
\begin{aligned}
  \tilde{t}(\vx)
  &=
  \operatorname{clip}
  \left(
    \frac{|\phi(\vx)|}{w_{\Gamma}},
    0,
    1
  \right),\\
  s_{\Gamma}(\vx)
  &=
  \tilde{t}(\vx)^2
  \left(
    3
    -
    2\tilde{t}(\vx)
  \right),
\end{aligned}
\label{eq:sdf-density-ramp}
\end{equation}
where $\operatorname{clip}(a,0,1)$ limits $a$ to the interval $[0,1]$. The
function $s_{\Gamma}$ rises smoothly from zero on $\Gamma$ to one at a distance
$w_{\Gamma}$, and stays equal to one beyond it. The desired cell length scale is
\begin{equation}
  \mu_{\Gamma}(\vx)
  =
  \mu_{\min}
  +
  \left(
    \mu_{\max}
    -
    \mu_{\min}
  \right)
  s_{\Gamma}(\vx),
\label{eq:sdf-adaptive-density}
\end{equation}
and the density is $\rho_{\Gamma}=\mu_{\Gamma}^{-4}$, as in
Eq.~\eqref{eq:density-sizing-relationship-2d}. We use $\mu_{\min}=0.03$,
$\mu_{\max}=0.06$, and $w_{\Gamma}=0.24$. With $\rho_{\Gamma}=1$, this term
is the CVT energy for a uniform density. Since $\rho_{\Gamma}$ is of order
$\mu_{\min}^{-4}\approx10^6$, the weight $\lambda_{\mathrm{in}}$ of this term
is much smaller than the other weights.

\textbf{The boundary cells.} The last two terms improve the shapes of the
boundary cells. Let $A_j$ and $P_j$ be the area and perimeter of the boundary
cell $j$. The area term
\begin{equation}
  \mathcal{L}_{\mathrm{area}}
  =
  \frac{1}{|\mathcal{B}_{\Gamma}|}
  \sum_{j\in\mathcal{B}_{\Gamma}}
  \left(
    \frac{A_j-\bar{A}_{\Gamma}}
    {\bar{A}_{\Gamma}+\varepsilon}
  \right)^2
\label{eq:sdf-boundary-quality-loss-area}
\end{equation}
penalizes differences between the areas of the boundary cells, where
\begin{equation}
  \bar{A}_{\Gamma}
  =
  \frac{1}{|\mathcal{B}_{\Gamma}|}
  \sum_{j\in\mathcal{B}_{\Gamma}}
  A_j
  \label{eq:sdf-mean-boundary-area}
\end{equation}
is their mean area. The circularity term
\begin{equation}
  \mathcal{L}_{\mathrm{circ}}
  =
  \frac{1}{|\mathcal{B}_{\Gamma}|}
  \sum_{j\in\mathcal{B}_{\Gamma}}
  (1-q_j)^2,
  \qquad
  q_j
  =
  \operatorname{clip}
  \left(
    \frac{4\pi A_j}{P_j^2+\varepsilon},
    0,
    1
  \right),
\label{eq:sdf-boundary-quality-loss-circ}
\end{equation}
makes the boundary cells rounder. The circularity $q_j$ is equal to one for a
circle, and is smaller for shapes that are less round. In both terms,
$\varepsilon$ is a small constant that avoids division by
zero; we use $\varepsilon=10^{-12}$.

All of these quantities are computed from the vertices of the cells, so their
gradient with respect to the generator positions is found by automatic
differentiation.

\subsubsection{Shape meshing examples}

\begin{figure}[!htpb]
  \centering
  \includegraphics[width=1\textwidth]
  {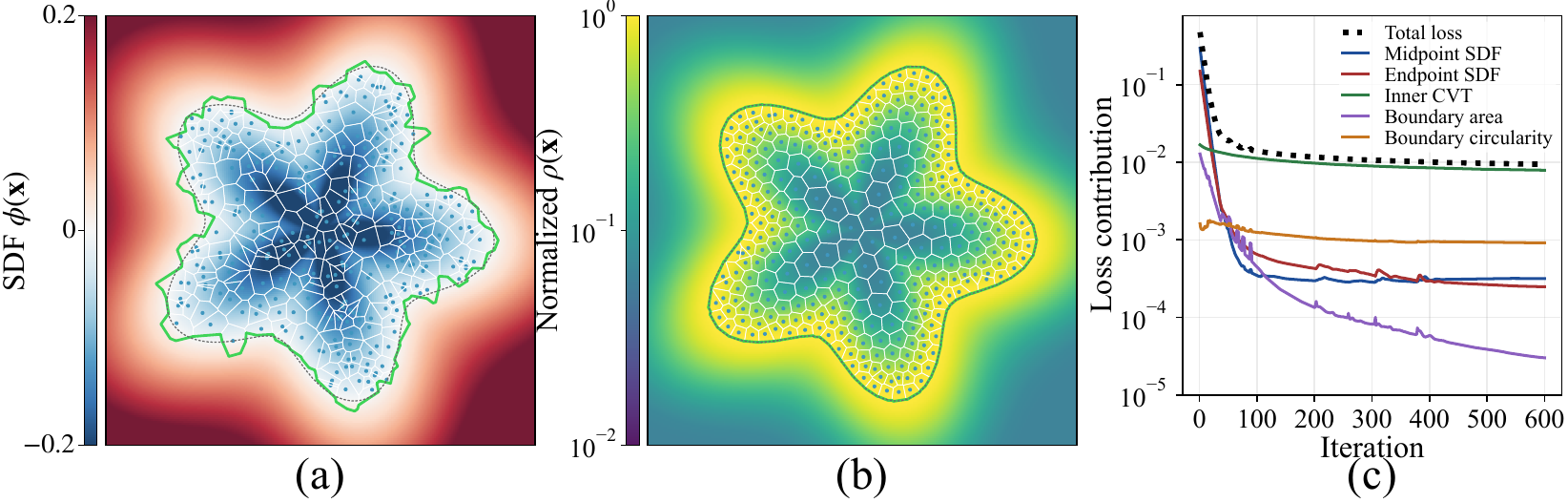} \caption{Meshing a
flower-shaped domain. (a)~Initial interface for random generator positions, over
the SDF $\phi$. (b)~Interface after 600 iterations of the Adam optimizer, over
the density $\rho_{\Gamma}$, normalized and shown on a logarithmic scale. The
gray dashed curve is the boundary $\Gamma$, green segments are the interface
edges, and blue points are the inner generators. (c)~The objective function and
its weighted terms during the optimization, with weights
$\lambda_{\mathrm{mid}}=10$, $\lambda_{\mathrm{end}}=2$,
$\lambda_{\mathrm{in}}=5\times10^{-4}$, $\lambda_{\mathrm{area}}=0.05$, and
$\lambda_{\mathrm{circ}}=0.02$.}
  \label{fig:sdf-flower-meshing}
\end{figure}

We tested this method on six shapes: a circle, a flower, a rounded square, a
heart, a star, and a peanut. The circle is smooth and convex, the flower and the
peanut have smooth concave parts, the rounded square has nearly flat sides and
rounded corners, and the heart and the star have sharp corners. In each case,
the shape is given only by its SDF. All six examples start from the same
$N=1000$ random generator positions, sampled independently and uniformly from
$[0.05,0.95]^2$ using random seed 13. The generators are optimized for 600
iterations using Adam with a learning rate of $5\times10^{-4}$. The weights
listed in the caption of Fig.~\ref{fig:sdf-flower-meshing} are used for all six
shapes.

Figure~\ref{fig:sdf-flower-meshing} shows the optimization for the flower. The
generators start at random positions, so the initial interface is far from
$\Gamma$. After 600 iterations, the interface follows $\Gamma$ closely, with
small cells next to $\Gamma$ and larger cells inside the shape. The objective
decreases rapidly during the first 100 iterations, and then slowly.

\begin{figure}[h]
  \centering
  \includegraphics[width=1\textwidth]
  {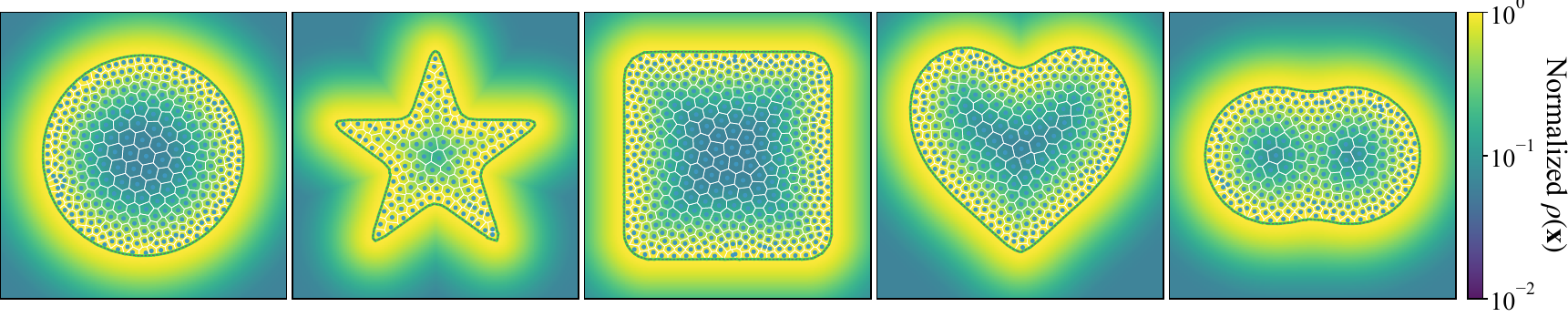}
\caption{Meshes of five shapes after 600 iterations of the Adam optimizer:
from left to right, a circle, a star, a rounded square, a heart, and a peanut.
Each panel shows the density $\rho_{\Gamma}$, normalized and on a logarithmic
scale. The gray dashed curve is the boundary $\Gamma$, green segments are the
interface edges, and blue points are the inner generators.}
  \label{fig:sdf-density-gallery}
\end{figure}

Figure~\ref{fig:sdf-density-gallery} shows the results for the other five
shapes after 600 iterations. In each case, the interface follows $\Gamma$
closely, and the cells are small along $\Gamma$ and larger away from it.

\FloatBarrier

\subsection{Anisotropic Voronoi foam design}
\label{sec:foam}

\begin{figure}[!htbp]
  \centering
      \includegraphics[width=\textwidth]
    {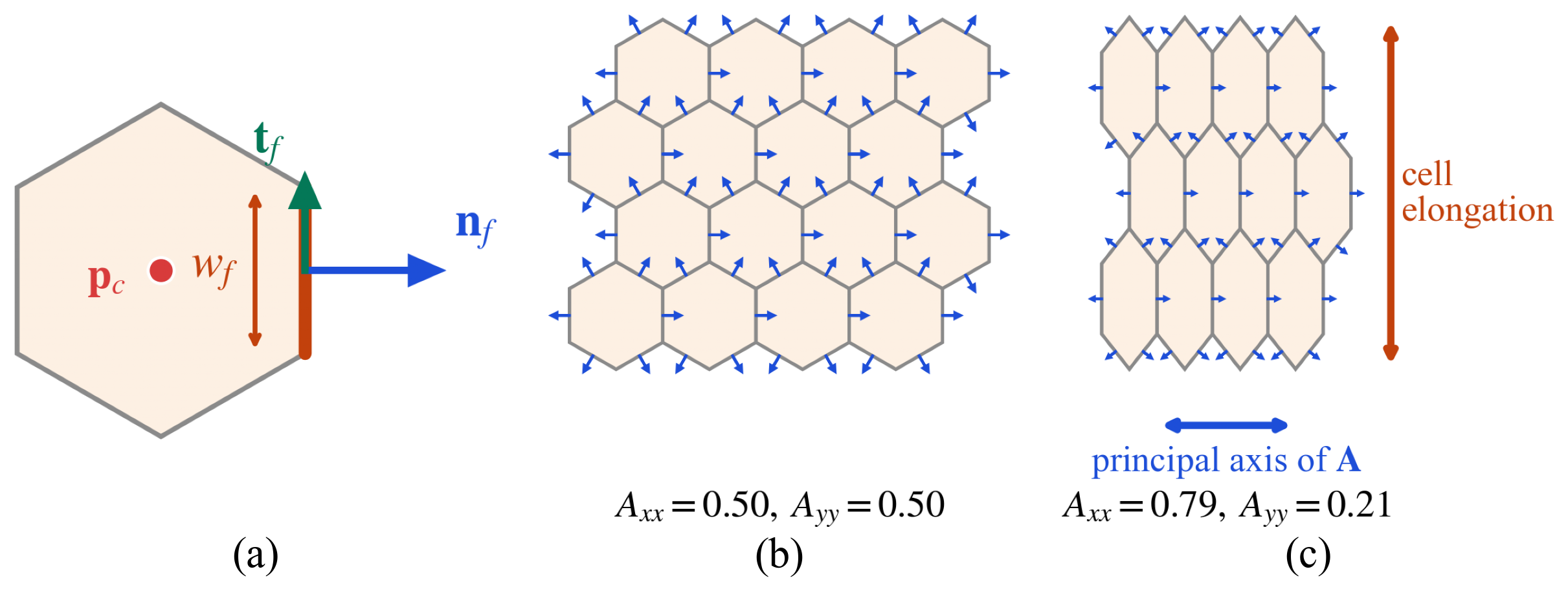}

  \caption{%
    Directions associated with the walls of a Voronoi cell. (a)~A cell with
    generator $\vect{p}_c$. For the highlighted wall $f$, $\vect{n}_f$ is the
    outward unit normal, $\vect{t}_f$ is a direction within the wall, and the
    length of the wall is its weight $w_f$ in Eq.~\eqref{eq:fabric-tensor}.
    (b)~Cells with the same size in all directions, for which the wall normals
    are spread evenly over all directions and
    $\vect{A}=\tfrac{1}{2}\vect{I}$. (c)~Cells elongated along $y$. The
    long walls are vertical, so their normals are mostly along $x$, and
    ${A}_{xx}>{A}_{yy}$. The principal axis of $\vect{A}$ (blue)
    is perpendicular to the direction in which the cells are elongated (orange).
    The tensors in (b) and (c) are computed from the cells that are drawn. Each
    interior wall is shared by two cells, so in (b) and (c) it has two arrows,
    one for the outward normal of each cell. These point in opposite
    directions, and give the same $\vect{n}_f\vect{n}_f^{\trans}$.%
  }
  \label{fig:foam-schematic}
\end{figure}

Foams whose cell walls have a preferred orientation arise in several settings.
Additively manufactured lattices can use oriented cells to tailor their
stiffness while keeping a low density~\cite{schaedler2016architected}. More
generally, the mechanical properties of a cellular solid depend strongly on
the geometry and topology of its cells, as well as on the material that it is
made from~\cite{gibson1997cellular}. Trabecular bone is a biological example:
its trabeculae tend to align with the directions in which the bone is
loaded~\cite{zysset2003review}.

The orientation of the walls is an important source of mechanical anisotropy.
Fabric tensors describe the directional structure of a cellular material, and
constitutive models can relate them to its elasticity
tensor~\cite{cowin1985relationship}. Under suitable assumptions, the principal
directions of the fabric tensor are the principal directions of the material.
Controlling the fabric tensor is therefore a geometric way to influence the
directional mechanical response of a foam. Our goal is to construct a Voronoi
foam whose fabric tensor matches a prescribed target.

Several directions are associated with a Voronoi cell, as shown in
Fig.~\ref{fig:foam-schematic}(a): the outward normal $\vect{n}_f$ of each wall
$f$, the directions within each wall, which are perpendicular to
$\vect{n}_f$, and the direction in which the cell is elongated. We use the
fabric tensor of Kanatani~\cite{kanatani1984}, which is defined by the
distribution of the wall normals. For a cell $c$ with walls
$\mathcal{F}(c)$, it is
\begin{equation}
  \vect{A}_c
  =
  \frac{\sum_{f\in\mathcal{F}(c)} w_f\,\vect{n}_f\vect{n}_f^{\trans}}
       {\sum_{f\in\mathcal{F}(c)} w_f},
  \label{eq:fabric-tensor}
\end{equation}
where the weight $w_f$ is the size of the wall: its length in 2D and its area
in 3D. Walls on the boundary of the domain are left out, since they are caused
by cutting the foam to the size of the domain, and are not part of its
structure. In 2D, with $\vect{n}_f=(n_{f,x},n_{f,y})$,
\begin{equation}
  \vect{A}_c
  =
  \frac{1}{\sum_f w_f}
  \sum_{f}
  w_f
  \begin{bmatrix}
    n_{f,x}^2        & n_{f,x}n_{f,y} \\[2pt]
    n_{f,x}n_{f,y}   & n_{f,y}^2
  \end{bmatrix}
  =
  \begin{bmatrix}
    A_{xx} & A_{xy} \\[2pt]
    A_{xy} & A_{yy}
  \end{bmatrix} .
  \label{eq:fabric-entries-2d}
\end{equation}
Since each $\vect{n}_f$ is a unit vector, $\operatorname{tr}(\vect{A}_c)=1$.
The diagonal entries measure how much the wall normals point along each
coordinate axis, and the off-diagonal entry, together with the diagonal
entries, determines the directions of the principal axes. If one direction
$\vect{u}$ dominates the fabric tensor, the wall normals tend to point along
$\vect{u}$, so the walls tend to be perpendicular to $\vect{u}$, and the cells
tend to be elongated perpendicular to $\vect{u}$. Since the tensor is computed
for each cell separately, the orientation can be controlled locally.

Each cell $c$ is given a target fabric tensor, built from a unit axis
$\vect{u}(c)$ and a strength $\alpha$,
\begin{equation}
  \vect{A}_{\mathrm{tgt}}(c)
  =
  \alpha\,\vect{u}(c)\vect{u}(c)^{\trans}
  +
  \frac{1-\alpha}{d-1}\bigl(\vect{I}-\vect{u}(c)\vect{u}(c)^{\trans}\bigr),
  \label{eq:target-tensor}
\end{equation}
where $d$ is the dimension. This tensor has the eigenvalue $\alpha$ along
$\vect{u}$, and the eigenvalue $(1-\alpha)/(d-1)$ in every direction
perpendicular to $\vect{u}$, so it has unit trace and treats all of the
directions perpendicular to $\vect{u}$ equally. In 2D, with
$\vect{u}=(\cos\varphi,\sin\varphi)$,
\begin{equation}
  \vect{A}_{\mathrm{tgt}}
  =
  \begin{bmatrix}
    \alpha\cos^2\!\varphi+(1-\alpha)\sin^2\!\varphi
      & (2\alpha-1)\cos\varphi\sin\varphi \\[4pt]
    (2\alpha-1)\cos\varphi\sin\varphi
      & \alpha\sin^2\!\varphi+(1-\alpha)\cos^2\!\varphi
  \end{bmatrix},
  \label{eq:target-entries-2d}
\end{equation}
whose off-diagonal entry is zero when $\vect{u}$ is along a coordinate axis or
when $\alpha=1/2$. For $\varphi=0$, the tensor is
$\operatorname{diag}(\alpha,1-\alpha)$, and in 3D, $\vect{u}=(1,0,0)$ gives
$\operatorname{diag}(\alpha,\tfrac{1-\alpha}{2},\tfrac{1-\alpha}{2})$.

The value $\alpha=1/d$ gives the isotropic tensor
$\vect{A}_{\mathrm{tgt}}=\vect{I}/d$. Values $\alpha>1/d$ favor wall normals
along $\vect{u}$, and so walls perpendicular to $\vect{u}$, which gives
plate-like (lamellar) cells. Values $\alpha<1/d$ favor walls parallel to
$\vect{u}$, which gives cells elongated along $\vect{u}$, in columns. Since
$\vect{u}$ can vary from cell to cell, the same formulation can be used for a
uniform foam orientation or for a foam with spatially varying orientations.

The objective function combines the mismatch in the fabric tensor with a
penalty on differences in cell size,
\begin{equation}
  \mathcal{L}
  =
  \mathcal{L}_{\mathrm{main}}
  +
  \lambda_s
  \frac{\operatorname{Var}(s_c)}{\bar{s}^2},
  \qquad
  \mathcal{L}_{\mathrm{main}}
  =
  \frac{1}{N}\sum_c
      \bigl\lVert \vect{A}_c-\vect{A}_{\mathrm{tgt}}(c) \bigr\rVert_F^2,
  \label{eq:foam-total-loss}
\end{equation}
where $N$ is the number of cells, $s_c$ is the size of cell $c$ (its area in 2D
and its volume in 3D), $\bar{s}$ is the mean cell size, and $\lambda_s=0.30$.
The second term is the squared coefficient of variation of the cell sizes, which
does not depend on the units of length. The first term,
$\mathcal{L}_{\mathrm{main}}$, controls the orientation of the walls but not the
sizes of the cells, and on its own, it can produce cells with very different
sizes. The second term prevents this without favoring any direction. In our
tests, these two terms were enough to give cells with the desired orientation
and similar sizes. The domain is the unit square in 2D and the unit cube in 3D.
The generators are initialized at independent uniform random positions in
$[0.05,0.95]^d$, so that none starts on the boundary, and are optimized with
Adam~\cite{kingma2014adam} with an initial learning rate of $4\times10^{-3}$ in
2D and $3\times10^{-3}$ in 3D, decreasing to one hundredth of that on a cosine
schedule. All examples are run with 1000 iterations.

To measure how well an optimized foam matches its target, we use the
orientation of each cell. Let $\vect{e}_c$ be the principal axis of
$\vect{A}_c$ that the target
distinguishes: the eigenvector with the largest eigenvalue when $\alpha>1/d$,
where the target concentrates wall normals along $\vect{u}(c)$, and the
eigenvector with the smallest eigenvalue when $\alpha<1/d$, where it starves
that direction of them. We then report
\begin{equation}
  M = \frac{1}{N}\sum_{c=1}^{N}\bigl(\vect{e}_c\cdot\vect{u}(c)\bigr)^2 ,
  \label{eq:alignment}
\end{equation}
which lies between 0 and 1. It equals 1 when every cell is aligned with its own
target axis, and decreases with misalignment. Because $M$ depends only on the
direction of each cell's principal axis, and not on how strongly the cell is
anisotropic, we also report the mean of
$\vect{u}(c)^{\trans}\vect{A}_c\,\vect{u}(c)$. This is the average of
$(\vect{n}_f\cdot\vect{u}(c))^2$ over the walls of each cell, weighted by their
size, and it approaches $\alpha$ when the target is met. Because the
tessellation changes discontinuously as the generators move, rounding
differences between computers grow during an optimization, and can change the
last digit of the values reported below.

\subsubsection{Two-dimensional example}
\label{sec:foam-2d}

\begin{figure}[!htbp]
  \centering
    \includegraphics[width=\textwidth]
    {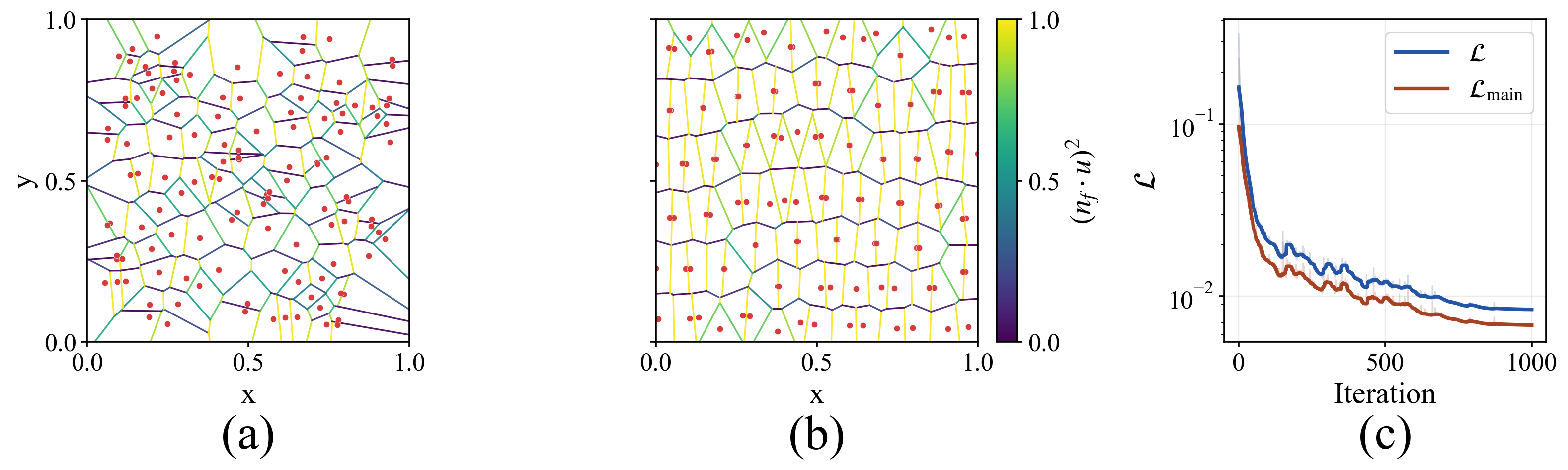}

  \caption{%
    Two-dimensional foam optimization with $N=120$ generators, constant axis
    $\vect{u}=(1,0)$, and strength $\alpha=0.75$. (a)~Initial foam.
    (b)~Optimized foam. In both panels each wall is colored by
    $(\vect{n}_f\cdot\vect{u})^2$, so that walls whose normal is along
    $\vect{u}$ are bright. Panels (a) and
    (b) use the same axes, and the color bar beside (b) applies to both.
    (c)~The objective function $\mathcal{L}$ and its first term
    $\mathcal{L}_{\mathrm{main}}$, the mismatch in the fabric tensor, during the
    optimization, with the raw values faint and their running median in bold.
  }
  \label{fig:foam-2d}
\end{figure}

We optimize a foam with $N=120$ generators, a constant axis $\vect{u}=(1,0)$,
and strength $\alpha=0.75$, so that
$\vect{A}_{\mathrm{tgt}}=\operatorname{diag}(0.75,0.25)$. Starting from random
generator positions, the alignment of Eq.~\eqref{eq:alignment} rises from
$M=0.454$ to $M=0.995$, and the mean of $\vect{u}^{\trans}\vect{A}_c\vect{u}$
rises from $0.481$ to $0.726$, close to the target strength $\alpha=0.75$. As
expected, the wall normals are mostly horizontal, so the walls are mostly
vertical and the cells are elongated vertically (Fig.~\ref{fig:foam-2d}). The
walls whose normal is within $45^{\circ}$ of the $x$ axis make up $74\%$ of the
total wall length, and the cells have a mean ratio of height to width of $2.2$.

\FloatBarrier

\subsubsection{Three-dimensional examples}
\label{sec:foam-3d}

In 3D, the method is the same, except that wall lengths are replaced by face
areas, and cell areas by cell volumes. We consider three targets, each with
$N=800$ generators, which differ only in the axis $\vect{u}(c)$ and the
strength $\alpha$ in Eq.~\eqref{eq:target-tensor}.

\begin{figure}[!htbp]
  \centering
    \includegraphics[width=\textwidth]
    {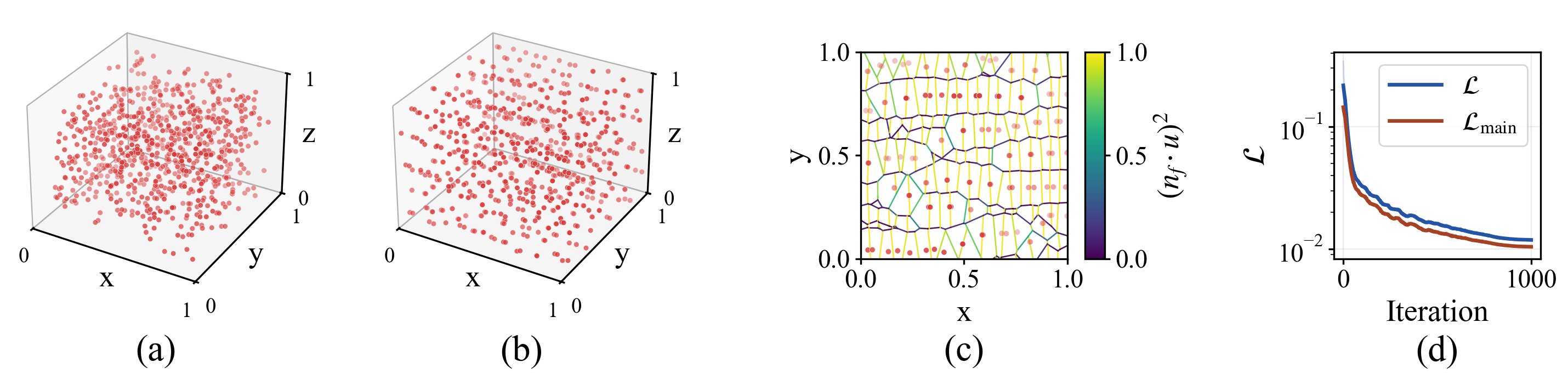}

  \caption{%
    Three-dimensional foam optimization with $N=800$ generators and the lamellar
    target, with the constant axis $\vect{u}=(1,0,0)$ and strength
    $\alpha=0.65$. (a)~Initial generator positions. (b)~Optimized generator
    positions. (c)~A slice through the optimized foam at $z=0.5$, with each wall
    colored by $(\vect{n}_f\cdot\vect{u})^2$, with the color bar beside it.
    (d)~The objective function
    $\mathcal{L}$ and its first term $\mathcal{L}_{\mathrm{main}}$ during the
    optimization, with the raw values faint and their running median in bold.
  }
  \label{fig:foam-3d-headline}
\end{figure}

The first target is lamellar, with the constant axis $\vect{u}(c)=(1,0,0)$ and
$\alpha=0.65$. Since $\alpha>1/3$, it favors wall normals along $x$, and so
walls perpendicular to $x$, giving plate-like cells stacked along $x$. This
case is shown in Fig.~\ref{fig:foam-3d-headline}.

The second target is columnar, with the constant axis $\vect{u}(c)=(0,0,1)$ and
$\alpha=0$, so that $\vect{A}_{\mathrm{tgt}}=\operatorname{diag}(0.5,0.5,0)$.
This target favors walls parallel to the $z$ axis, and treats the $x$ and $y$
directions equally. The cross-sections of the cells are nearly round, with a
median aspect ratio of $1.25$, where the aspect ratio of a cell is the square
root of the ratio of the two eigenvalues of the covariance matrix of its
vertices, projected onto the $xy$ plane. The $zz$ component of $\vect{A}_c$,
averaged over the cells, is $0.121$ rather than its target of zero, which would
require every interior wall to be exactly parallel to the $z$ axis, and so every
generator to lie at the same height. The size penalty is not the cause:
repeating the run with $\lambda_s=0$, from the same initial generators and for
the same 1000 iterations, lowers this average only to $0.092$. A more likely
explanation is that reaching this configuration from random positions needs a
large, coordinated rearrangement of the generators, which gradient descent does
not find.

The third target has an axis that varies in space, pointing horizontally away
from the vertical center line of the domain,
\begin{equation}
  \vect{u}(c)
  =
  \frac{(p_x-\tfrac12,\;p_y-\tfrac12,\;0)}
       {\sqrt{(p_x-\tfrac12)^2+(p_y-\tfrac12)^2}} ,
  \label{eq:radial-field}
\end{equation}
where $\vect{p}_c=(p_x,p_y,p_z)$ is the generator of cell $c$, with
$\alpha=0.65$ as in the lamellar case. This favors walls that face the center
line, which arranges the cells in curved slats around it.

Table~\ref{tab:foam-alignment} reports $M$ before and after optimization for the
three targets. In all three cases, it rises from about $1/3$, the value for
cells whose axes are unrelated to $\vect{u}$, to above $0.99$, so almost every
cell is aligned with its own target axis. The table also reports the mean of
$\vect{u}(c)^{\trans}\vect{A}_c\,\vect{u}(c)$, which approaches the target
strength $\alpha$ in the two cases with $\alpha=0.65$. The columnar target, with
$\alpha=0$, is the exception, and is discussed above.

\begin{table}[h]
\centering
\begin{tabular}{lrrrrr}
\toprule
 & \multicolumn{2}{c}{$M$}
   & \multicolumn{2}{c}{Mean $\vect{u}^{\trans}\vect{A}_c\vect{u}$} & \\
\cmidrule(lr){2-3} \cmidrule(lr){4-5}
Target & Initial & Optimized & Initial & Optimized & $\alpha$ \\
\midrule
Lamellar & 0.322 & 0.994 & 0.328 & 0.607 & 0.65 \\
Columnar & 0.343 & 0.992 & 0.336 & 0.121 & 0    \\
Radial   & 0.308 & 0.993 & 0.315 & 0.605 & 0.65 \\
\bottomrule
\end{tabular}
\caption{The alignment $M$ of Eq.~\eqref{eq:alignment} and the mean of
$\vect{u}(c)^{\trans}\vect{A}_c\,\vect{u}(c)$ for the three three-dimensional
targets, before and after optimization. The second quantity approaches the
target strength $\alpha$.}
\label{tab:foam-alignment}
\end{table}

\begin{figure}[!htbp]
    \centering
    \includegraphics[width=\textwidth]
    {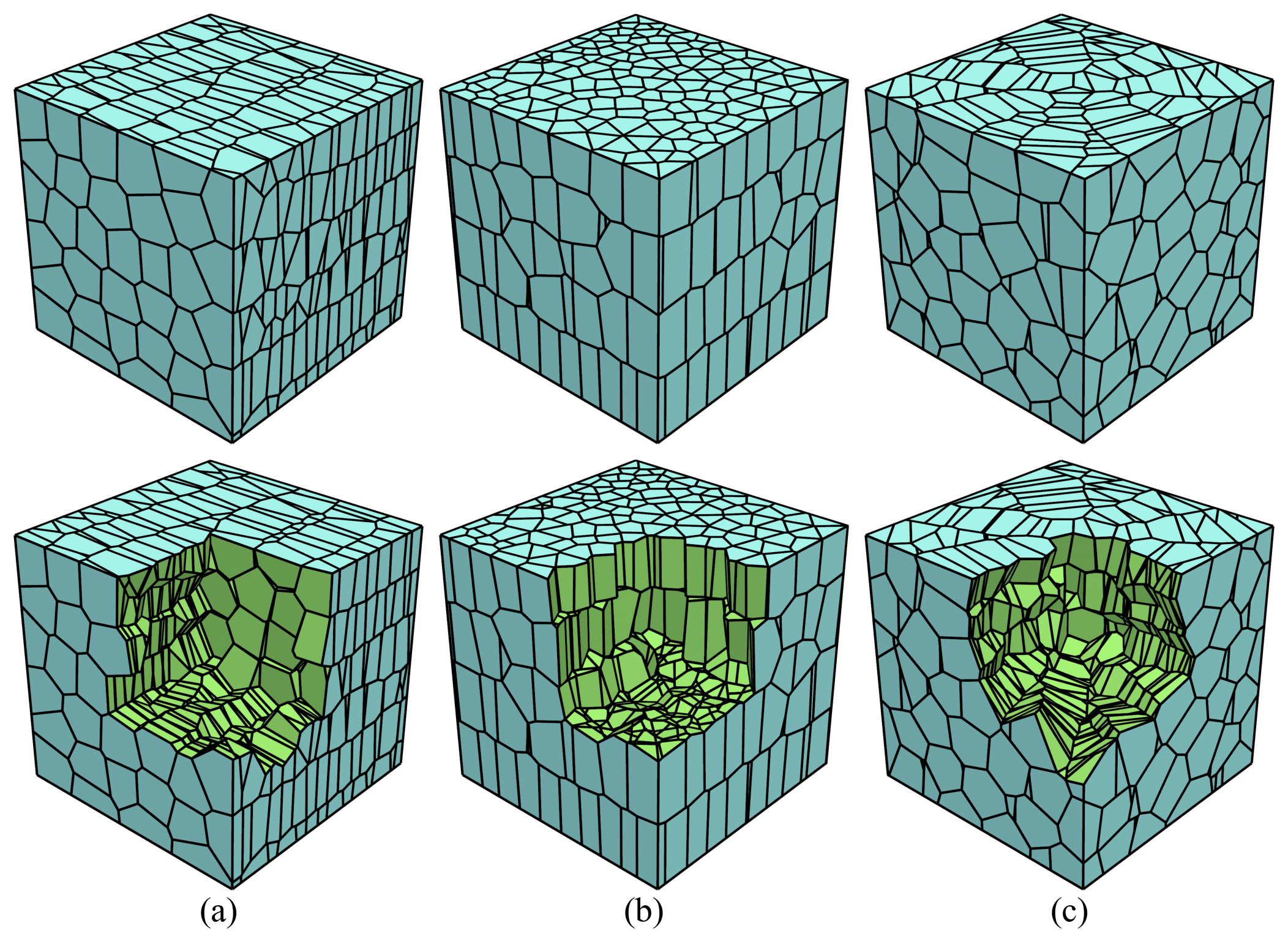}

    \caption{%
      Three-dimensional foam optimization with $N=800$ generators:
      (a)~lamellar target, $\vect{u}=(1,0,0)$ with $\alpha=0.65$;
      (b)~columnar target, $\vect{u}=(0,0,1)$ with $\alpha=0$;
      (c)~radial axis of Eq.~\eqref{eq:radial-field}, with $\alpha=0.65$.
      The top row shows the whole foams, and the bottom row shows them with the
      cells in one corner removed, so that the cells inside can be seen.%
    }
    \label{fig:foam-3d-grid}
\end{figure}

Figure~\ref{fig:foam-3d-grid} shows the three optimized foams, drawn with
PyVista~\cite{sullivan19} from meshes exported by \vplex. In the bottom row, the
cells in one corner of the cube are removed, so that the cells inside can be
seen. Whole cells are removed, so every surface that is shown is a complete
Voronoi wall. The walls on the boundary of the cube are light cyan, and the
exposed interior walls are light green. The lamellar target gives plates stacked
along $x$, the columnar target gives columns along $z$ with nearly round
cross-sections, and the radial target gives curved slats around the vertical
center line.

\FloatBarrier

\section{Conclusion and future work}
\label{sec:conclusion}

We introduced \vplex, an open-source Python package built on the multi-threaded
\vpp{} library, which computes Voronoi tessellations in parallel and makes them
differentiable. It offers three representations of the result. The geometry
representation, which is the default, gives the vertices, faces, and neighbors
of each cell, as computed by \vpp. The per-cell representation also labels each
vertex with the generators that define it. \vplex{} uses these labels to compute
the derivative of each vertex with respect to the generator positions, so that
PyTorch can find the gradient of any quantity computed from the cells. The
topology representation is used to build differentiable meshes in which each
vertex is shared between cells. \vplex{} can also compute radical tessellations
of particles with different radii, and their derivatives with respect to the
radii. It also provides tools for extracting the interface between two groups of
cells, analyzing cells, and exporting results.

\vplex{} was faster than the other Python libraries that we tested. For $10^6$
generators on 28 threads, it is 46 times faster than SciPy in 2D and 39 times
faster in 3D. The \vpp{} cell computation scales well with the number of
threads, with adjusted efficiencies of 0.96 in 3D and 0.91 in 2D on 28 threads.
The main limits on the parallel efficiency are a small number of steps that run
on one thread, creating the \vpp{} container and freeing memory, and the copy of
the result into NumPy arrays. In a CVT optimization on the CPU, most of the time
is spent in PyTorch, evaluating the energy and its derivatives, while on the GPU
the tessellation takes the largest share of the time. We demonstrated the
package with three applications: CVTs with a variable density, meshing of 2D
shapes defined by signed distance functions, and the design of foams with a
preferred cell orientation.

There are several directions for future work. The steps that run on one
thread could be made faster by allocating the memory of the \vpp{} container in
parallel, or avoided in an optimization by reusing the container from one
iteration to the next. On the GPU, the derivative of the vertices is currently
computed on the CPU, which requires copying data between the GPU and the CPU;
computing it on the GPU would remove this cost. Computing the tessellation
itself on the GPU would remove the largest remaining cost of an optimization
on the GPU, although this would be a substantial project.

\section*{Acknowledgements}
J.L.~was partially supported by the UCLA Institute for Digital Research and
Education (IDRE) Postdoc Fellowship and by an Association for Women in
Mathematics (AWM) Mentor Travel Grant. C.H.R.~was partially supported by the
Applied Mathematics Program of the U.S.~DOE Office of Science Advanced
Scientific Computing Research under Contract No.~DE-AC02-05CH11231. We also
acknowledge support from the Toyota Research Institute and NVIDIA.

\section*{Declaration of generative AI and AI-assisted technologies in the
  writing process}

During the preparation of this work the authors used Claude (Anthropic) and
Codex (OpenAI) in order to assist with drafting and editing the text, and with
writing and documenting code. After using these tools, the authors reviewed
and edited the content as needed and take full responsibility for the content
of the publication.

\bibliography{voroplex}

\end{document}